\documentclass[aps,pra,twocolumn,amsmath,amssymb,preprintnumbers,floatfix]{revtex4-1}

\usepackage[utf8]{inputenc}
\usepackage{txfonts}
\usepackage{microtype}
\usepackage{eucal}
\usepackage{bm}
\usepackage{physics}
\usepackage{xr}

\usepackage{color}
\usepackage{soul}

\usepackage{graphicx}
\usepackage{float}

\usepackage[colorlinks,allcolors=blue]{hyperref}
\usepackage[capitalize]{cleveref}

\setcitestyle{numbers,square}

\newcommand{\anticomm}[2]{\{{#1}\,,\,{#2}\}}                    

\newcommand{\eg}{\textit{e.g.}\ }

\newcommand{\ve}[1]{\boldsymbol{#1}}

\newcommand{\x}{\lambda}  

\newcommand{\zh}{\ve{\hat{z}}}

\newcommand{\SO}{{\rm SO}}
\let\nodag\mathstrut

\newcommand*{\citen}[1]{%
  \begingroup
    \romannumeral-`\x 
    \setcitestyle{numbers,square}%
    \cite{#1}%
  \endgroup   
}

\usepackage{soul}

\newif\ifcolourjacob
\colourjacobtrue
\newif\ifcolourhaakon
\colourhaakontrue
\begin{document}
\title{Tunable superconducting critical temperature in ballistic hybrid structures \\
with strong spin-orbit coupling}
\author{Haakon T. Simensen and Jacob Linder}
\affiliation{Center for Quantum Spintronics, Department of Physics, Norwegian University of Science and Technology, NO-7491 Trondheim, Norway
}

\date{\today}
 
\begin{abstract}
\noindent
We present a theoretical description and numerical simulations of the superconducting transition in hybrid structures including strong spin-orbit interactions. The spin-orbit coupling is taken to be of Rashba type for concreteness, and we allow for an arbitrary magnitude of the spin-orbit strength as well as an arbitrary thickness of the spin-orbit coupled layer. This allows us to make contact with the experimentally relevant case of enhanced interfacial spin-orbit coupling via atomically thin heavy metal layers. We consider both interfacial spin-orbit coupling induced by inversion asymmetry in an S/F-junction, as well as in-plane spin-orbit coupling in the ferromagnetic region of an S/F/S- and an S/F-structure. Both the pair amplitudes, local density of states and critical temperature show dependency on the Rashba strength and, importantly, the orientation of the exchange field. In general, spin-orbit coupling increases the critical temperature of a proximity system where a magnetic field is present, and enhances the superconducting gap in the density of states. We perform a theoretical derivation which explains these results by the appearance of long-ranged singlet correlations. Our results suggest that $T_{\rm c}$ in ballistic spin-orbit coupled superconducting structures may be tuned by using only a single ferromagnetic layer.
\end{abstract}

\maketitle

\section{Introduction}

In recent years, different ways to exert spin-control over the superconducting state and properties has garnered increasing interest \cite{linder_nphys_15, beckmann_review}. This includes phenomena such as spin-polarized supercurrents \cite{Robinson2010, khaire_prl_10} where, traditionally, magnetic inhomogeneities have played a key role in this endeavour as they provide a source to spin-polarized Cooper pairs \cite{bergeret_prl_01, bergeret_rmp_05, Eschrig2011}. However, more recently the focus has shifted to exploiting spin-orbit interactions as a way to 
achieve a spin-dependent coupling to the superconducting state. Effects such as magnetoanisotropic
supercurrents \cite{jacobsen_scirep_16, costa_prb_17, hogl_prl_15}, anisotropic and paramagnetic Meissner effects \cite{espedal_prl_16}, thermospin effects \cite{bobkova_prb_17a, bobkova_prb_17b}, and 
spin-galvanic couplings \cite{konschelle_prb_15, amundsen_prb_17} have very recently been investigated in this context. We note in particular
that a recent experiment \cite{banerjee_arxiv_17} reported a spin-valve effect on the superconducting transition temperature
$T_c$ in a layered Nb/Pt/Co/Pt structure. This contrasts previous superconducting spin-valve measurements
where two ferromagnets were used \cite{gu_prl_02, moraru_prl_06, leksin_prl_12} instead of a single magnetic layer. The spin-valve effect is made possible due to the 
thin Pt layers which provide Rashba spin-orbit interactions due to interfacial inversion symmetry
breaking. 

Motivated by this experiment and the interesting physics arising in spin-orbit coupled hybrid structures including superconducting elements, we here present a study of the critical temperature, local density of states, and the induced pairing correlations in such systems. We use a fully quantum mechanical treatment and solve the BdG-equations in the ballistic limit. With a non-zero exchange field in the ferromagnetic region, both the pair amplitude, local density of states and critical temperature show dependency on the strength and, importantly, the orientation of the exchange field. In general, spin-orbit coupling increases the critical temperature of the system, and strengthens the superconducting gap in the density of states. We also present results for the same observable quantities for in-plane spin-orbit coupling in the ferromagnetic region of an S/F/S- and an S/F-structure. The results are similar to interfacial spin-orbit coupling, although the effect is in general stronger. Additionally, this type of spin-orbit coupling gives rise to a stronger anisotropy in the dependence on the exchange field direction. Our results demonstrate how $T_c$ may be controlled with a single ferromagnetic layer in ballistic spin-orbit coupled superconducting hybrids.

\section{Theory and methods}
\subsection{Spin-orbit coupling}\label{ssec:SOC}

A commonly used model for the spin-orbit Hamiltonian in systems where structural inversion asymmetry is broken, for instance by interfaces, is the Rashba Hamiltonian given by \cite{Bychkov1984,Manchon2015}

\begin{equation}
H_\SO = \alpha_{\rm R} \left( \hat{\ve{n}} \times \boldsymbol{\hat{\sigma}} \right) \cdotp \ve{k},
\label{eq:H_SO_nonH}
\end{equation}

\noindent where $\alpha_{\rm R}$ is the Rashba parameter, $\hat{\ve{n}}$ is the unit vector pointing in the direction of the broken inversion symmetry, and $\boldsymbol{\hat{\sigma}}$ is the vector of Pauli matrices. We will refer to  $\ve{\tilde{h}}_\SO \equiv \alpha_{R} \left( \hat{\ve{n}} \times \ve{k} \right)$ as the SOC-induced field. To ensure that we use a Hermitian Hamiltonian, we symmetrize it by letting $\alpha_{\rm R}(x) k_x \rightarrow \frac{1}{2}\anticomm{\alpha_{\rm R}(x)}{k_x}$, and the Hamiltonian thus becomes $H_\SO = \frac{1}{2} \anticomm{\alpha_{\rm R}}{\ve{k}} \cdotp \left( \hat{\ve{n}} \times \boldsymbol{\hat{\sigma}} \right),$ where $\{\ldots\}$ is an anticommutator. This procedure is necessary in hybrid structures, as considered in this paper, where SOC exists only in certain layers. In order to test the physical validity of this Hamiltonian in an actual system, one could for instance do spin-resolved ARPES measurements to test how the crystal momentum of the electrons correlate with their spin orientation.

\subsection{Psuedospin and Cooper pairs}\label{ssec:pseudospin}

Let us briefly explain how Cooper pairs are formed in systems with both SOC and a magnetic field present. For simplicity, we start by defining a Hamiltonian in which the magnetic field is perpendicular to the SOC-induced fields. Let the magnetic field be $\ve{h} = h_0 \zh$, and restrict the SOC-induced field to be parallel to the $x$-axis and proportional to $k_y$. This type of SOC may be physically realized by for instance interfacial SOC between two regions in a two-dimensional system spanning the $yz$-plane. By writing out the Pauli matrices explicitly, the Hamiltonian follows as

\begin{equation}
\mathbf{H} =
-
\begin{pmatrix}
h_0 & \alpha_{\rm R} \hat{k}_y \\
\alpha_{\rm R} \hat{k}_y & -h_0 \\
\end{pmatrix}
.
\label{eq:PauliSO}
\end{equation}

\noindent As the Hamiltonian includes SOC, spin is no longer conserved, and we refer to pseudospin, $\sigma'$, as the new well-defined quantum number. It can be shown that the singlet $s$-wave Cooper pair projects onto the new eigenstate basis as

\begin{equation}
\begin{split}
\ket{k_y, \uparrow}&\ket{-k_y,\downarrow} - \ket{k_y, \downarrow} \ket{-k_y,\uparrow} \\
&= \cos\left(\theta_{\rm SO}\right) \Bigg[ \ket{k_y, \uparrow^{'}}\ket{-k_y,\downarrow^{'}} - \ket{k_y, \downarrow^{'}} \ket{-k_y,\uparrow^{'}} \Bigg] \\
&-
\sin\left(\theta_{\rm SO}\right) \Bigg[ \ket{k_y, \uparrow^{'}}\ket{-k_y,\uparrow^{'}} + \ket{k_y, \downarrow^{'}} \ket{-k_y,\downarrow^{'}} \Bigg]
\end{split}
.
\label{eq:singlet}
\end{equation}

\noindent where $\cos(\theta_\SO) = \frac{h_0}{\tilde{h}}$, $\sin(\theta_\SO) = \frac{\alpha_{\rm R} k_y}{\tilde{h}}$, and {$\tilde{h} = \sqrt{h_0^2 + \alpha_{\rm R}^2 |k_y|^2}$}. In the presence of only a magnetic field, the singlet state does not transform at all since we in this limit have $\sigma=\sigma'$ and $\theta_\text{SO}=0$. With SOC present however, it is evident that the singlet state projected onto the eigenbasis results in both a pseudospin-singlet and a pseudospin-triplet component.

Since Cooper pairs are comprised of electrons of approximately equal energy, Eq. \eqref{eq:singlet} must be modified in order to reflect the real pairings. We have to pair electrons of different momenta, such that the momentum shift cancels the energy difference caused by the SOC and the magnetic field. As pseudospin by definition defines the two possible eigenstates of the Hamiltonian in Eq. \eqref{eq:PauliSO} for a given momentum $k_y$, it is apparent that electrons with equal pseudospin and $|k_y|$ are found at the same energy level, while the ones with opposite pseudospin and equal $|k_y|$ are found at different energy levels. We thus treat pseudospin just as we treat spin with only magnetic fields present. That is, we define a shifted momentum $k_y^{\pm} = k_y + (\Delta k)^{\pm}$, where the $\pm$ applies for pseudospin up/down. $(\Delta k)^{\pm}$ is defined such that the different single-particle pseudospin states involved in the two-particle states have equal energy. By using the notation $\ket{k_y^{\pm},\sigma'} = \ket{k_y,\sigma'} e^{i (\Delta k)^{\pm} x}$, we can express the $s$-wave singlet Cooper pair wave function as 

\begin{equation}
\begin{split}
\psi_{\perp}(x) &\sim \cos\left(\theta_{\rm SO}\right) \Bigg\{ \ket{k_y^{+}, \uparrow^{'}}\ket{-k_y^{-},\downarrow^{'}} e^{i [ (\Delta k)^{+} - (\Delta k)^{-} ] x} \\
&\quad\quad\quad\quad\quad -
\ket{k_y^{-}, \downarrow^{'}} \ket{-k_y^{+},\uparrow^{'}} e^{-i[ (\Delta k)^{+} - (\Delta k)^{-} ] x} \Bigg\} \\
&-
\sin\left(\theta_{\rm SO}\right) \Bigg\{ \ket{k_y^{+}, \uparrow^{'}}\ket{-k_y^{+},\uparrow^{'}} + \ket{k_y^{-}, \downarrow^{'}} \ket{-k_y^{-},\downarrow^{'}} \Bigg\}
\end{split}
,
\label{eq:psi_SOC_perp}
\end{equation}

\noindent where we have neglected the change in $\theta_{\rm SO}$ due to momentum shift, and where the $\perp$ indicates that the magnetic field and SOC-induced fields are orthogonal to each other. One may observe that the pseudospin-singlet component of the singlet state gains a phase shift, whereas the pseudospin-triplet component does not. We choose to adapt the terminology which is frequently used on spin-triplet pairs, and name the pseudospin-triplet pair which is not subject to a pair-breaking phase a {\it long-ranged} pair. From this analysis, it is evident that a fraction of the $s$-wave Cooper singlet pair can adapt a long-ranged behaviour in a system featuring both Rashba SOC and a magnetic exchange field. It therefore follows that the singlet pair in such a system is partly long-ranged and partly short-ranged. In contrast, in the absence of SOC, the $s$-wave singlet pair would have adapted only short-ranged behaviour. 

If we define the system such that the magnetic field and SOC-induced fields are parallel, spin would still be a good (conserved) quantum number. As a consequence, no pseudospin-triplet component of the singlet state will appear, and the singlet state remains purely short-ranged. If we had made the analysis completely general, that is include all intermediate angles, the algebra would have become messy. However, the analysis would have revealed that the transition between orthogonal and parallel setup happens smoothly and gradually, and the parallel and orthogonal setup therefore represents two extrema. Another consequence of an arbitrary magnetization alignment is that both the singlet states and all the triplet states would have been projected onto the eigenbasis as linear combinations of both the pseudospin-singlet and all the pseudospin-triplets. A consequence of this is that mixing between the Cooper pair spin-states may occur.

\subsection{Solving the BdG equations}

The Hamiltonian for a system which includes a magnetic exchange field $\ve{h}$, and Rashba SOC reads \cite{Fossheim2004,Halterman,Bychkov1984}

\begin{equation}
\resizebox{1.0\hsize}{!}{$
\begin{split}
H &= \sum_{\sigma} \int d\ve{r} \hat{\psi}^{\dagger}(\ve{r},\sigma) \Big[ H_e - \ve{h}(\ve{r}) \cdotp \boldsymbol{\hat{\sigma}} + i \alpha_{\rm R}(\ve{r}) (\boldsymbol{\hat{\sigma}} \times \hat{\ve{n}}) \cdotp \nabla  \Big] \hat{\psi}^{\nodag}(\ve{r},\sigma) 
\\
&\quad 
+ \int d\ve{r} \bigg\{ \Delta^{*}(\ve{r}) \hat{\psi}^{\nodag}(\ve{r},\downarrow) \hat{\psi}^{\nodag}(\ve{r},\uparrow) + \Delta^{\nodag}(\ve{r}) \hat{\psi}^{\dagger}(\ve{r},\uparrow) \hat{\psi}^{\dagger}(\ve{r},\downarrow)  \bigg\}.
\label{eq:H}
\end{split}$}
,
\end{equation}

\noindent where $\hat{\psi}^{\dagger}$ and $\hat{\psi}^{\nodag}$ are electronic creation and annihilation operators respectively, and $H_e = \frac{\hbar^2 \ve{k}^2}{2m} + V(\ve{r})$, that is the combined kinetic energy and non-magnetic potential energy. We have assumed that the SOC-term of the Hamiltonian is Hermitian. If this is not the case, the symmetrization procedure presented in the last section must be applied. We will now specialize this Hamiltonian to a two-dimensional system spanning the $xy$-plane. We follow closely the technical procedure presented in many papers by K. Halterman and various coauthors, for instance in Ref. \citen{Halterman2015}. The system is translationally invariant in the $y$-direction, and of length $d$ in the $x$-direction. 

We first perform a Bogoliubov transformation of the electronic operators,

\begin{align}
\begin{split}
\hat{\psi}^{\nodag}(x,\uparrow) &= \sum_{n} \bigg[ u_{n,\uparrow}(x) \gamma_{n}^{\nodag} - v_{n,\uparrow}^*(x) \gamma_{n}^{\dagger} \bigg], \\
\hat{\psi}^{\nodag}(x,\downarrow) &= \sum_{n} \bigg[ u_{n,\downarrow}(x) \gamma_{n}^{\nodag} + v_{n,\downarrow}^*(x) \gamma_{n}^{\dagger} \bigg],
\end{split}
\label{eq:bt}
\end{align}

\noindent where $u_{n,\uparrow}(x)$ and $v_{n,\uparrow}(x)$ are quasielectron and quasihole wave functions, respectively. We have assigned a label $n$ to each state, denoting they are energy eigenstates. This operator transformation should by definition transform the Hamiltonian into the form ${H = \sum_{n,\sigma} E_{n,\sigma} \gamma_{n,\sigma}^{\dagger} \gamma_{n,\sigma}^{\nodag}}$. These quasiparticle amplitudes are found by solving the Bogoliubov-de Gennes (BdG) equations. It is however convenient to first expand these amplitudes as Fourier series. With $\psi_n \equiv [u_{n,\uparrow}, u_{n,\downarrow}, v_{n,\uparrow}, v_{n,\downarrow}]$, the Fourier expansion follows as

\begin{equation}
\psi_{n}(x) = \sqrt{\frac{2}{d}} \sum_{q = 1}^{\infty} \hat{\psi}_{nq} \sin(k_q x),
\label{eq:Fourier_psi}
\end{equation}

\noindent where $k_q = q \pi / d$, and where the components of $\hat{\psi}_{nq} = [\hat{u}_{nq}^{\uparrow}, \hat{u}_{nq}^{\downarrow} ,\hat{v}_{nq}^{\uparrow}, \hat{v}_{nq}^{\downarrow}]^{\rm T}$ are the Fourier components of the expansion. By expanding the wave functions in sine-functions, the boundary conditions are automatically satisfied. It is moreover useful to define new spin-orbit operators, $h^i_{\rm SO}$,

\begin{align}
\begin{split}
\tilde{h}^i_{\rm SO} u_{n,\sigma} &\equiv h^i_{\rm SO} u_{n,\sigma}, \\
\tilde{h}^i_{\rm SO} v_{n,\sigma} &\equiv -h^i_{\rm SO} v_{n,\sigma}.
\end{split}
\label{eq:SO_op}
\end{align}

\noindent where the particle/hole-dependence is isolated in the sign convention. With these definitions, the BdG equations in Fourier space are given by

\begin{widetext}
\begin{equation}
\begin{split}
\begin{pmatrix}
\hat{H}_e - \hat{h}^z -\hat{h}^z_{\rm SO}										& -\hat{h}^x + i \hat{h}^y -\hat{h}^x_{\rm SO} + i \hat{h}^y_{\rm SO}	& 0 																					& \hat{\Delta}^{\nodag} \\
-\hat{h}^x - i \hat{h}^y -\hat{h}^x_{\rm SO} - i \hat{h}^y_{\rm SO}		& \hat{H}_e + \hat{h}^z + \hat{h}^z_{\rm SO}									& \hat{\Delta}^{\nodag} 													& 0 \\
0 																																		& \hat{\Delta}^{*} 																										& -(\hat{H}^*_e - \hat{h}^z + \hat{h}^z_{\rm SO}) 						 & -\hat{h}^x - i \hat{h}^y +\hat{h}^x_{\rm SO} + i \hat{h}^y_{\rm SO}\\
\hat{\Delta}^{*} 																											& 0 																																	& -\hat{h}^x + i \hat{h}^y +\hat{h}^x_{\rm SO} - i \hat{h}^y_{\rm SO}	& -(\hat{H}^*_e + \hat{h}^z - \hat{h}^z_{\rm SO})
\end{pmatrix}
\begin{pmatrix}
\hat{u}_{n}^{\uparrow} \\
\hat{u}_{n}^{\downarrow} \\
\hat{v}_{n}^{\uparrow} \\
\hat{v}_{n}^{\downarrow}
\end{pmatrix}
= E_{n}
\begin{pmatrix}
\hat{u}_{n}^{\uparrow} \\
\hat{u}_{n}^{\downarrow} \\
\hat{v}_{n}^{\uparrow} \\
\hat{v}_{n}^{\downarrow}
\end{pmatrix}
,
\end{split}
\label{eq:BdG_full}
\end{equation}
\end{widetext}

\noindent where we have defined $\hat{u}_n^{\sigma} = [\hat{u}_{n1}^{\sigma}, \hat{u}_{n2}^{\sigma}, \hat{u}_{n3}^{\sigma}, ...]$ and $\hat{v}_n^{\sigma} = [\hat{v}_{n1}^{\sigma}, \hat{v}_{n2}^{\sigma}, \hat{v}_{n3}^{\sigma}, ...]$. The BdG equations determine the quasiparticle amplitudes, as well as the energy spectrum. The matrix elements appearing above are defined as

\begin{align*}
\hat{H}_{e}(q,q') &= \frac{2}{d} \int_{0}^{d} dx \; \sin(k_{q'} x) \Bigg[ \frac{\hbar^2}{2 m} \bigg( \frac{ \pi q}{d} \bigg)^2 + V(x) + E_{\perp} - E_{\rm F} \Bigg] \\
&\quad \times \sin(k_{q} x),
\end{align*}
\begin{align*}
\hat{\Delta}(q,q') &= \frac{2}{d} \int_{0}^{d} dx \; \sin(k_{q'} x) \Delta(x) \sin(k_{q} x),
\end{align*}
\begin{align*}
\hat{h}_i(q,q') &= \frac{2}{d} \int_{0}^{d} dx \; \sin(k_{q'} x) h_i(x)  \sin(k_{q} x),
\end{align*}
\begin{align*}
\hat{h}^{i}_{\rm SO}(q,q') &= \frac{2}{d} \int_{0}^{d} dx \; \sin(k_{q'} x) h^{i}_{\rm SO}(x)  \sin(k_{q} x),
\end{align*}

\noindent where $i \in \{x,y,z\}$ in the two last definitions.

One of the main goals of solving the BdG equations is finding the superconducting energy gap, $\Delta$. It is defined as 

\begin{equation}
\Delta(\ve{r}) = V_{\rm SC}(x) \expval{\hat{\psi}^{\nodag}(x,\uparrow) \hat{\psi}^{\nodag}(x,\downarrow)},
\label{eq:s_wave}
\end{equation}

\noindent where $V_{\rm SC}$ is a coupling strength between electrons inside the energy interval ${[E_{\rm F} - \hbar \omega_{\rm D}, \, E_{\rm F} + \hbar \omega_{\rm D}]}$. By insertion of the Bogoliubov transformation in Eq. \eqref{eq:bt}, and ${V_{\rm SC}(x) = \lambda(x) / D_2(E_{\rm F})}$, where the weak-coupling constant $\lambda$ is finite inside superconductors and zero elsewhere, while ${D_2(E_{\rm F}) = \frac{m}{\pi \hbar^2}}$ is the energy-independent density of states per area in two dimensions, we obtain

\begin{equation}
\resizebox{1.0\hsize}{!}{$
\Delta(x) = \lambda(x) \frac{E_{\rm F}}{4 k_{\rm F}} \sum_{n}^{'} [u_{n,\uparrow}(x) v_{n,\downarrow}^*(x) + u_{n,\uparrow}(x) v_{n,\downarrow}^*(x)] \tanh(E_{n}/2k_{\rm B}T),
$}
\label{eq:Gap}
\end{equation}

\noindent The sum over $n$ is a sum over all energy eigenstates, which formally is a sum over all eigenstates of Eq. \eqref{eq:BdG_full} for every possible value of $E_{\perp}$. The primed summation indicates that this is a constrained sum over energy levels within the energy interval where $s$-wave singlet Cooper pairing occurs. This opens up for using a self-consistent approach. We start out by {\it guessing} an initial $\Delta$. The closer the initial guess is to the actual $\Delta$, the fewer iterations through the BdG equations are necessary. Before starting this procedure, we make $\Delta$ a dimensionless quantity by letting ${\Delta(x)/\Delta_0 \rightarrow \Delta(x)}$, where $\Delta_0$ is the bulk value of the superconducting energy gap within a clean superconductor. $\Delta(x)$ should therefore presumably be constrained to $|\Delta(x)| \le 1$. For most situations, using a zeroth order approach by guessing $\Delta = 1$ inside superconducting regions, and $\Delta = 0$ elsewhere, is a sufficiently accurate starting point. Solve the BdG equations in \eqref{eq:BdG_full} for this $\Delta$, and obtain a set of eigenvectors $\psi_n$. Use this set of eigenvectors to define a new $\Delta$ using Eq. \eqref{eq:Gap}, and repeat this procedure until $\Delta$ converges towards the true superconducting gap. In this paper, we stopped the procedure when $\Delta$ at no point had a relative change of more than $10^{-3}$ between two consecutive iterations.

\subsection{Pair amplitudes}

The singlet energy gap captures the singlet correlation within the superconducting regions of a system. Outside of these regions, the amplitude is by definition identically zero due to $\lambda(x)$ being zero. To provide information on the proximity effect, that is how far into non-superconducting regions superconducting order penetrates, we define the $s$-wave singlet pair amplitude

\begin{equation}
f_0(x) = \frac{\Delta(x)}{\lambda(x)}.
\label{eq:singlet_amp}
\end{equation}

\noindent This amplitude is chosen to be normalized to $|f_0| \le 1$. We furthermore define the $s$-wave triplet amplitudes, that is the odd-frequency triplets, as \cite{Halterman2008}

\begin{align}
f_1(x,\tau) &= \frac{1}{2} \expval{\hat{\psi}^{\nodag}(x,\tau,\uparrow) \hat{\psi}^{\nodag}(x,0,\downarrow) + \hat{\psi}^{\nodag}(x,\tau,\downarrow) \hat{\psi}^{\nodag}(x,0,\uparrow)}, \\
f_2(x,\tau) &= \frac{1}{2} \expval{\hat{\psi}^{\nodag}(x,\tau,\uparrow) \hat{\psi}^{\nodag}(x,0,\uparrow) - \hat{\psi}^{\nodag}(x,\tau,\downarrow) \hat{\psi}^{\nodag}(x,0,\downarrow)}, \\
f_3(x,\tau) &= \frac{1}{2} \expval{\hat{\psi}^{\nodag}(x,\tau,\uparrow) \hat{\psi}^{\nodag}(x,0,\uparrow) + \hat{\psi}^{\nodag}(x,\tau,\downarrow) \hat{\psi}^{\nodag}(x,0,\downarrow)}, \\
\end{align}

\noindent where $\tau$ is the relative time coordinate. We name the triplets captured by the $f_1$-amplitude $(s_z = 0)$-pairs, reflecting that these have zero spin projection along the $z$-axis, thus being $\sigma_z$-eigenstates. We name the Cooper pairs responsible for the $f_2$- and $f_3$-amplitudes $(s_z = \pm 1)$-triplets. We underline that these are not $\sigma_z$-eigenstates, and hence have no well defined $s_z$ as they are linear combinations of two-particle states with $s_z = \pm 1$. By insertion of the Bogoliubov transformations in Eq. \eqref{eq:bt}, and utilizing the identities which were used in the derivation of Eq. \eqref{eq:Gap}, we obtain

\begin{align}
f_{1}(x,\tau) &= \frac{1}{2} \sum_{n} \Big[ u_{n,\uparrow}(x) v^{*}_{n,\downarrow}(x) - u_{n,\downarrow}(x) v^{*}_{n,\uparrow}(x) \Big] \zeta_n(\tau), \\
f_{2}(x,\tau) &= -\frac{1}{2} \sum_{n} \Big[ u_{n,\uparrow}(x) v^{*}_{n,\uparrow}(x) + u_{n,\downarrow}(x) v^{*}_{n,\downarrow}(x) \Big] \zeta_n(\tau),\\
f_{3}(x,\tau) &= -\frac{1}{2} \sum_{n} \Big[ u_{n,\uparrow}(x) v^{*}_{n,\uparrow}(x) - u_{n,\downarrow}(x) v^{*}_{n,\downarrow}(x) \Big] \zeta_n(\tau),
\end{align}

\noindent where ${\zeta_n(t) = \sin \Big(\frac{E_n \tau}{\hbar} \Big) - i \cos \Big( \frac{E_n \tau}{\hbar} \Big) \tanh \Big(\frac{E_n}{2k_{\rm B} T} \Big)}$. In this paper, the triplet pair amplitudes have been normalized with the same prefactor as $f_0$. Additionally, we only plot the real part of the pair amplitudes.

\subsection{LDOS}

The local density of states (LDOS), $N(E,x)$, provides information on the distribution of states as a function of energy and position. Its interpretation is that $N(E,x) dE$ equals the number of quantum states within the infinitesimal energy interval ${[E, E + dE]}$ at position $x$. It can be expressed as \cite{Halterman}

\begin{equation}
N(E,x) = \sum_n {\nodag} \sum_{\sigma} \Big\{ \abs{u_{n \sigma}(x)}^2 \delta(E - E_n) + \abs{v_{n \sigma}(x)}^2 \delta(E + E_n) \Big\},
\end{equation}

\noindent where the $\delta$-function is the Dirac delta function. As all the energy levels are discretized, $N(E,x)$ will be a discrete distribution function. To smoothen out the density of states, we perform a convolution with a Gaussian of width ${0.02 \Delta_0}$. In this paper, the LDOS is normalized to be 1 in the normal metal limit, that is several times $\Delta_0$ away from $E_{\rm F}$.

\subsection{Critical temperature}

For the calculation of the critical temperature, we follow closely the procedure of Ref. \citen{Barsic2007}, where $T_{\rm c}$ is found by treating $\Delta$ as a small first-order perturbation. We can then solve the BdG equations once to zeroth order, that is with $\Delta = 0$, and use perturbation theory to define a finite $\Delta$ from the eigenvectors. This first-order $\Delta$ will be $T$-dependent, and we find $T_{\rm c}$ by identifying the point where $\Delta = 0$ is the only possible solution. The complete derivation of this perturbative approach is performed in Appendix \ref{sec:App_Tc}. Such a procedure assumes that the superconducting transition is not a first order one, since in that case $\Delta$ cannot be made arbitrarily close to zero. The result is a matrix eigenvalue problem,

\begin{equation}
\Delta_l^{(1)} = \sum_k \mathbf{J}_{l k}(T) \Delta_k^{(1)},
\label{eq:T_c_eig}
\end{equation}

\noindent where the matrix elements $\textbf{J}_{l k}$ are defined by the formula

\begin{equation}
\begin{split}
\textbf{J}_{l k}(T) &= \lambda  \frac{2 E_{\rm F}}{k_{\rm F} d^3}  \sum_n \sum_{m_{\parallel}} \sum_{p,q} K_{pql} \\
&\Bigg\{ 
 \ve{v}_{mq}^{(0) \dagger} \textbf{J}_2 \ve{u}_{np}^{(0)} \frac{\sum_{i,j} \ve{u}_{ni}^{(0) \dagger} \textbf{J}_2 \ve{v}_{mj}^{(0)} K_{ijk} }{E_n^{p} - E_m^{h}} \tanh(\frac{E_n^{p}}{2 k_B T}) \\
&+
\ve{v}_{nq}^{(0) \dagger} \textbf{J}_2 \ve{u}_{mp}^{(0)} \frac{\sum_{i,j} \ve{u}_{mi}^{(0) \dagger} \textbf{J}_2 \ve{v}_{nj}^{(0)} K_{ijk} }{E_n^{h} - E_m^{p}} \tanh(\frac{E_n^{h}}{2 k_B T})
\Bigg\},
\end{split}
\label{eq:J_matrix}
\end{equation}

\noindent where $\ve{v_{nj}} = [v^{\uparrow}_{nj},v^{\downarrow}_{nj}]^{T}$ and $\ve{u}_{nj} = [u^{\uparrow}_{nj},u^{\downarrow}_{nj}]^{T}$ are vectors of quasiholes and quasielectrons, $(0)$-superscipt denotes zeroth order, $\mathbf{J}_2$ is the ${(2\times2)}$ exchange matrix (defined in Appendix \ref{sec:App_Tc}), and $E_n^{h}$ and $E_n^{p}$ are the zeroth order energy spectra of quasiholes and quasielectrons, respectively. To simplify notation, we have introduced ${K_{ijk} = \int_0^d dx \Theta(x-x_0) \sin(k_i x) \sin(k_j x) \sin(k_k x)}$. The sums over $i$, $j$, $p$ and $q$ go over the Fourier wave numbers. The constrained sum over $n$ goes over the kinetic energy contributions from all directions. The sum over $m_{\parallel}$ goes over all kinetic energy contributions from the $x$-direction, with $m_{\perp} = n_{\perp}$ implied.

Eq. \eqref{eq:T_c_eig} is a matrix eigenvalue equation. It has one obvious solution, the trivial solution, that is $\Delta(x) = 0$. This solution is of no particular interest, since it implies that superconductivity is absent. If we assume $\Delta(x) \ne 0$ however, the equation has a solution if and only if the matrix $\mathbf{J}(T)$ has an eigenvalue which is 1. Since superconductivity is sensitive to temperature, one should therefore expect only the trivial solution to remain if $T > T_{\rm c}$, where $T_c$ is the critical temperature where superconductivity breaks down. This involves that all the eigenvalues of $\mathbf{J}(T)$ falls below 1. The critical temperature is therefore found by identifying at which temperature the largest eigenvalue of $\mathbf{J}(T)$ drops below 1.

\section{Results and Discussion}
SOC lifts spin-rotational symmetry, and thus the simultaneous presence of SOC and a magnetic field should reveal spin-anisotropic behaviour of superconductivity. This is what motivates us to explore SOC in F/S-structures. We will look at two types of spin-orbit coupling. First, we will explore the highly localized interfacial SOC, of which results are given in Sec. \ref{ssec:IntSOC_FS}. We will thereafter look at in-plane SOC inside F-regions, of which results are given in Secs. \ref{ssec:InplaneSOC_SFS} and \ref{ssec:InplaneSOC_SF}.

We use $\hbar \omega_{\rm D} / E_{\rm F} = 0.04$ for all calculations, as well as $T = 0$ in all calculations for the pair amplitudes and LDOS. The exchange field strength is chosen up to $h_0/E_{\rm F} = 0.3$, most realistically realized by placing transition metal ferromagnets such as Fe, Ni, or Co in the ferromagnetic region. The Rashba strength is varied between up to $\alpha_{\rm R} k_{\rm F}/E_{\rm F} = 0.5$. Large SOC effects are probably easiest to realize at the surface of heavy metals such as Au or Pt, as it has been reported that these structures give a Rashba effect two orders of magnitude larger than in semiconducting 2DEGs \cite{Manchon2015}. As we in this paper would like to explore the general effects of combining exchange fields and the Rashba effect, we vary $\alpha_{\rm R}$ over a relatively wide range. The dimensions used in this paper coincide with the routinely achieved experimental dimensions in heterostructures.

In all systems, we assume that the Fermi level $E_{\rm F}$ is equal and constant in all regions. Furthermore, we assume that the effective masses, work functions and densities are equal in all regions, in addition to there being no scattering potential in the interface between the regions. This is clearly a crude simplification, and in order to provide a more a realistic description of specific materials one should rather use parameters obtained from experiments. In this paper, we have chosen not to include these parameters, as doing so would result in more undetermined parameters that would complicate the analysis. The main purpose of this paper is to show the \textit{qualitative} effect of combining exchange fields and SOC, and we therefore choose the simplest approximation, namely that all of these parameters are constant throughout the system. Despite the crude approximation, similar routines (excluding SOC) have previously provided results which coincide well with experimental results \cite{Halterman2002}. For future work, it is straightforward to include Fermi level mismatch and interface scattering potentials in the numerical method, and thereby obtain results which are closer to specific materials.


\subsection{Interfacial SOC in an F/S-structure}
\label{ssec:IntSOC_FS}

\begin{figure}[htb!]
\centering
\includegraphics[width=1.0\columnwidth]{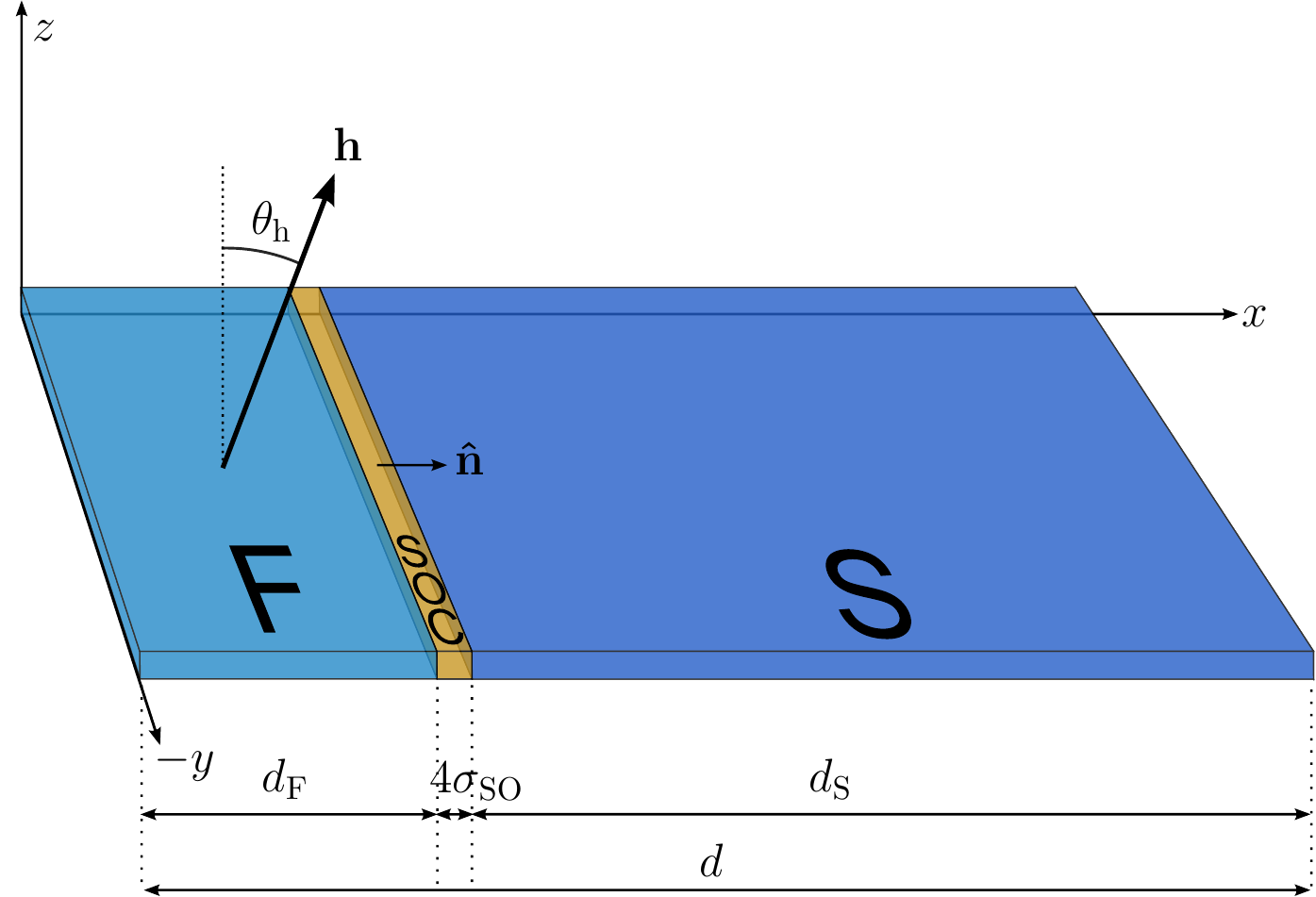}
\caption{An illustration of the F/S-structure with SOC in the junction. The system considered is in reality not of restricted length along the $y$-axis, but is of infinite extent in this direction. Moreover, the structure is of zero height, that is of no extent in the $z$-direction.}
\label{FsoS_sketch}
\end{figure}

We start by looking at interfacial SOC in an F/S-stucture. The system is two-dimensional, of length $d = 1.23\xi_0$ in the $x$-direction, and is translationally invariant in the $y$-direction. The length of the F-region has been set to $0.2\xi_0$, and the superconductor's length has been set to $\xi_0,$ where $\xi_0$ is the coherence length of the superconductor. In between these regions, there is a SOC-layer of width $0.03\xi_0$. The system is illustrated in Fig. \ref{FsoS_sketch}. The SOC-potential has been Gaussian distributed inside this region, that is $\alpha_{\rm R}(x) \sim N(x; \lambda_\SO, \sigma_\SO)$, where the expectation value of the distribution, $\lambda_\SO$, is in the middle of the SOC-region. The variance of the distribution is $\sigma_\SO^2$, and $4\sigma_\SO = 0.03\xi_0$ covers most of the distribution. We use a Rashba coupling strength of $\alpha_{\rm R} k_{\rm F} / E_{\rm F} = 0.5$. In the F-region, we define a magnetic field ${\ve{h} = h_0 \big( \sin\left(\theta_h\right) \ve{\hat{x}} + \cos\left(\theta_h\right) \ve{\hat{z}} \big)}$, in which we set $h_0 / E_{\rm F} = 0.3$.

\subsubsection{Pair amplitudes}

\begin{figure}[htb!]
\centering
\includegraphics[width=1.0\columnwidth]{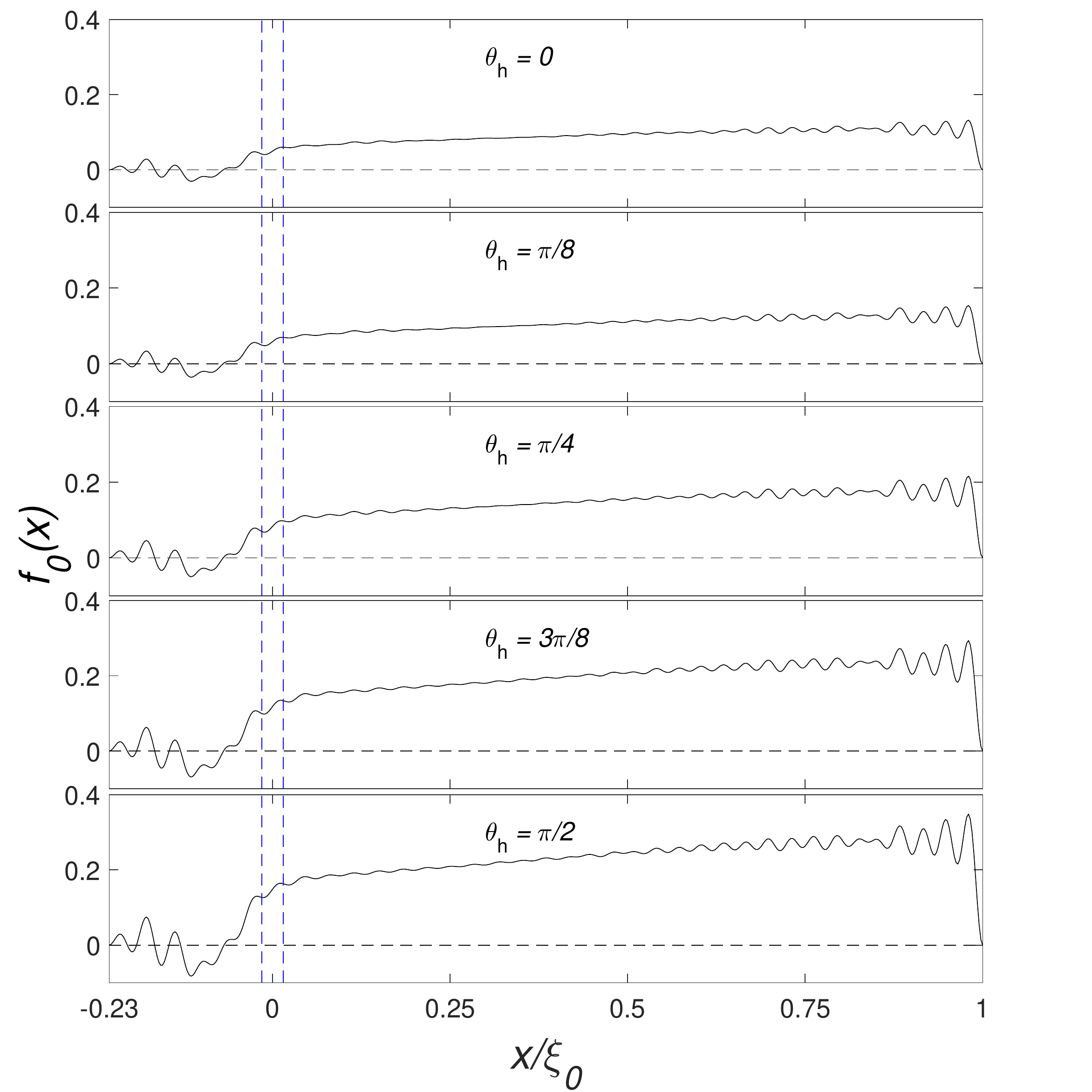}
\caption{The singlet pair amplitude plotted for five different magnetization angles, $\theta_h$, for an F/S-structure with SOC in a thin layer at the interface. The SOC-layer is Gaussian distributed within the blue dotted lines (which cover a width of $4\sigma_{\rm SO}$).}
\label{fig:f0_FSSOC}
\end{figure}

The $s$-wave singlet amplitude is plotted for five different magnetization angles, $\theta_h$, in Fig. \ref{fig:f0_FSSOC}.
The upper plot shows the results for $\theta_h = 0$, and the magnetization angle is increased by $\pi/8$ for every plot downwards. It can be observed that the singlet correlation, and thus the superconducting pair potential, grows by increasing the magnetization angle. At $\theta_h = 0$, its maximum before the oscillations at the boundary is approximately $0.1$. Growing steadily by increasing magnetization angle, this maximum doubles as $\theta_h$ approaches $\pi/2$. Hence, it seems as though a magnetization perpendicularly aligned to the SOC-induced fields results in best conditions for superconductivity to exist.

If the magnetic field and SOC-induced fields are aligned in the $z$-direction, $s_z$ is still a conserved quantum number. For the $s$-wave singlet Cooper pairs, the SOC then mimics a potential barrier, causing the F- and S-regions to be partly decoupled. In this case, we would therefore in general expect SOC to protect the superconducting state to some extent by damping the proximity effect. If the magnetic field and SOC-induced fields are not aligned however, we cannot precisely make qualitative predictions by evaluating $s_z$-states. We therefore turn to the pseudospin eigenstates, derived in Sec. \ref{ssec:pseudospin}. The main result of this section was that if the SOC-induced field is perpendicular to a magnetic field, a component of the singlet state becomes long-ranged. That is, if we project the singlet state onto the eigenbasis, it will in general be a linear combination of a pseudospin-singlet and a $(s' = \pm 1)$-pseudospin-triplet, the latter of which do not gain a relative phase throughout the system due to having zero CoM. As a consequence of this effect, the leakage of singlets is reduced, allowing for a larger singlet amplitude to sustain. This effect of SOC is $\theta_h$-dependent, and will therefore increase as $\theta_h$ increases. The results obtained by numerical calculations seem to support this analysis.

\begin{figure}[htb!]
\centering
\includegraphics[width=1.0\columnwidth]{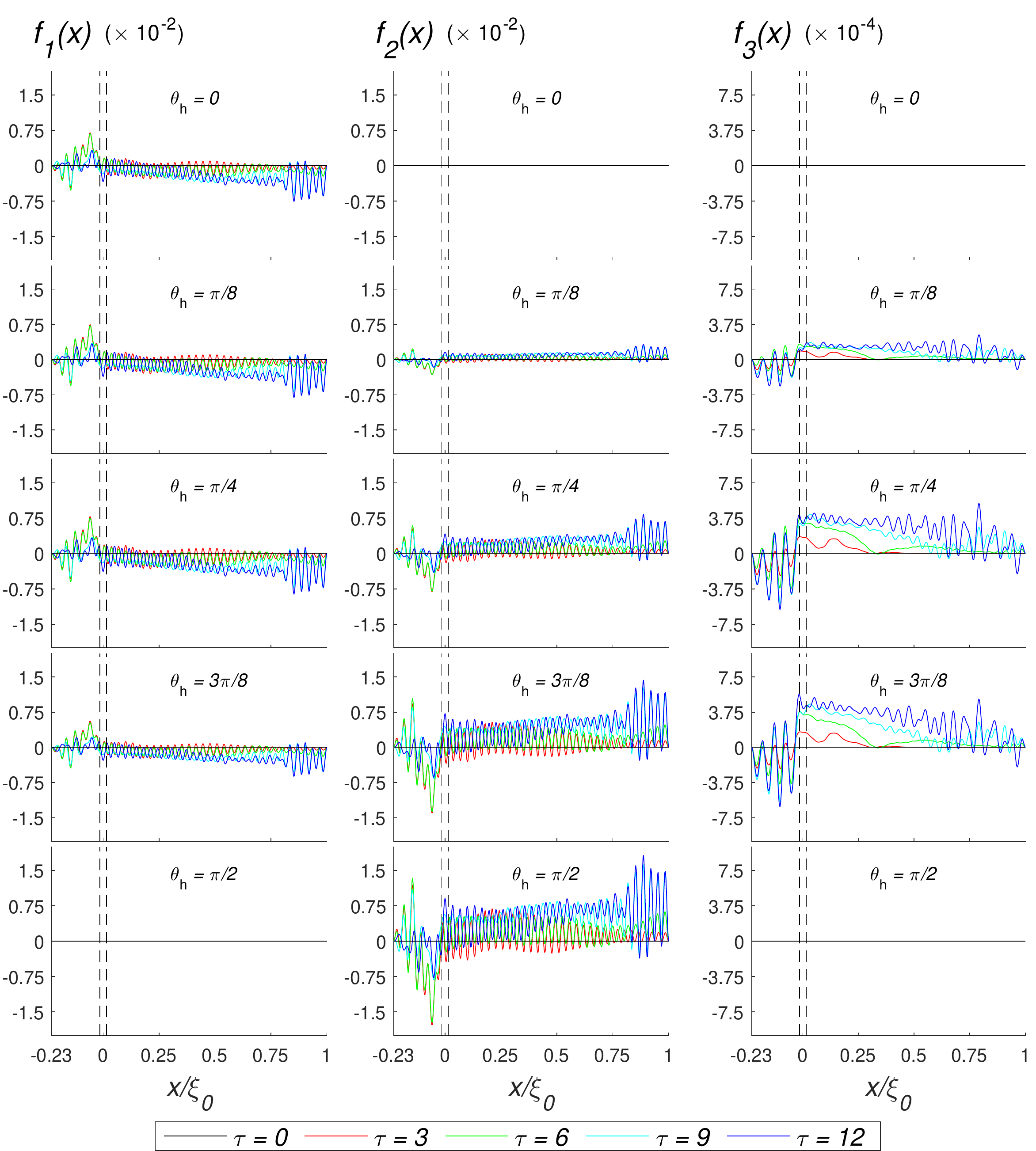}
\caption{The triplet amplitudes for five different magnetization angles, $\theta_h$, for the F/S-structure with a SOC-layer in the junction. Each plot contains triplet correlations for five different relative times, $\tau$.}
\label{fig:triplet_FSSOC}
\end{figure}

Another prediction from Sec. \ref{ssec:pseudospin} was that mixing between the triplet pairs should occur at intermediate angles, due to they using a common set of pseudospin channels through the system. This is verified by Fig. \ref{fig:triplet_FSSOC}, where all triplet amplitudes are plotted for five different relative times $\tau = \omega_{\rm D} t$.
In the absence of SOC, the $f_3$-amplitude would not have appeared by rotating the magnetic field in the $xz$-plane. With SOC however, it clearly appears, and this must therefore be due to Cooper pair spin-mixing caused by SOC.s At $\theta_h = 0$, $s_z$ is a conserved quantum number, and no $(s_z = \pm 1)$-amplitudes may be produced. For increasing magnetization angles, the spin-mixing effect seem to grow. At $\theta_h = \pi/2$ however, the shift of spin basis does not cause spin mixing, as predicted by the discussion in Sec. \ref{ssec:pseudospin}.


\subsubsection{LDOS}
\begin{figure}[htb!]
	\begin{center}
		\includegraphics[width=1.0\columnwidth]{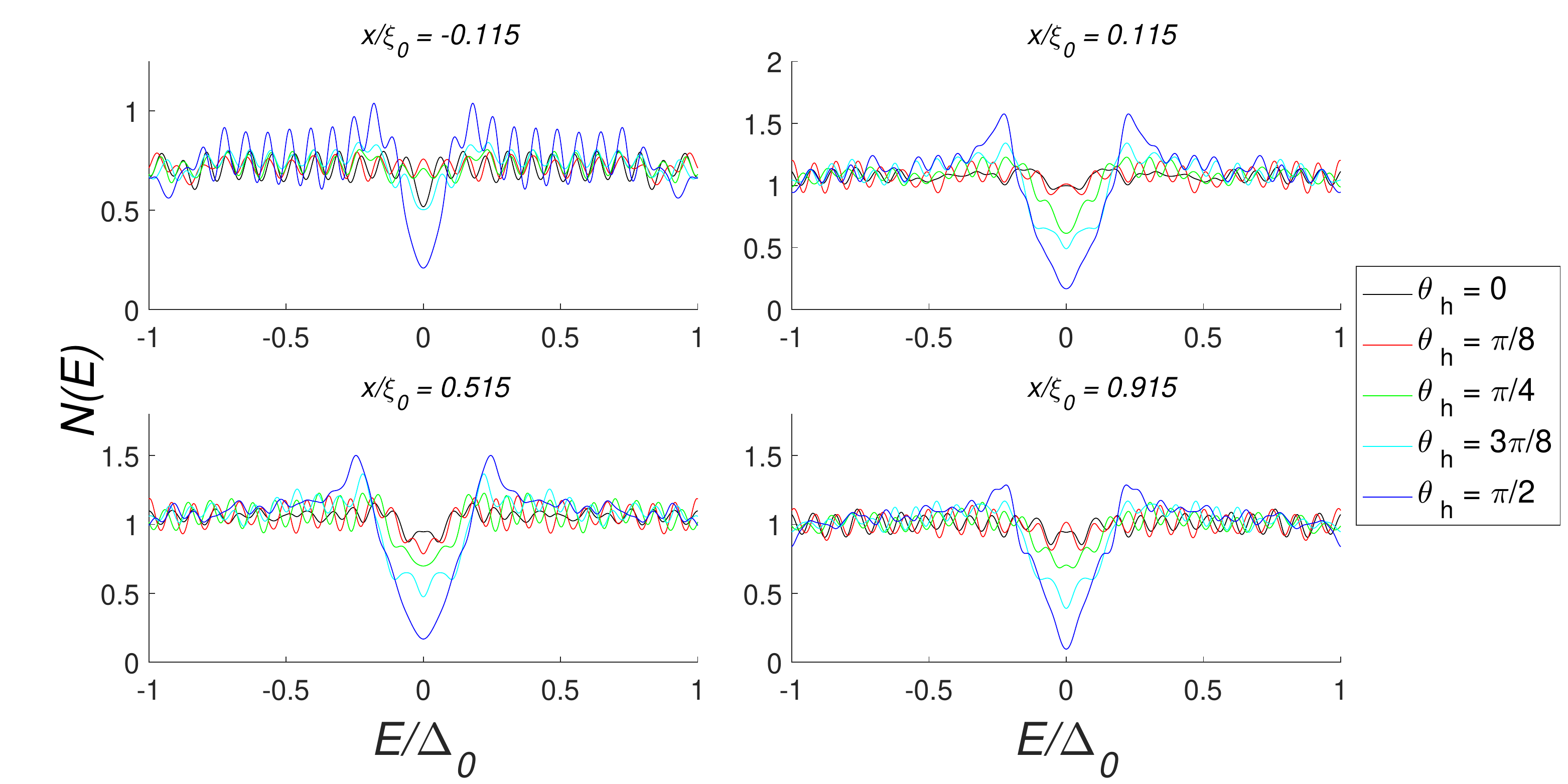}
	\end{center}
	\caption{The LDOS for the F/S-structure with SOC in the interface plotted at four different positions, as indicated above each plot. At each position, the LDOS is plotted for five different magnetization angles, $\theta_h$. The results are obtained with $N_{\perp}$ = 2000.}
	\label{fig:LDOS_FS_SO}
\end{figure}

As a consequence of the analysis so far, we expect the band gap to be more developed for higher magnetization angles, $\theta_h$. This is due to the creation of long-ranged singlets, which should imply fewer triplet states relative to singlet states, thus reducing the number of states within the band gap. When $\theta_h = 0$, this effect does not occur, and the plots should be qualitatively rather equal to a clean F/S-junction. For $\theta_h = \pi/2$, the effect should be at its maximum, creating the most prominent band gap. The LDOS at four different positions are plotted in Fig. \ref{fig:LDOS_FS_SO}, both inside the F- and S-region.

The plots show very clearly that the superconducting gap becomes much more prominent for higher magnetization angles. For $\theta_h = 0$, one can in fact almost not spot any gap at all. As we rotate $\theta_h$ further towards $\pi/2$, this gap grows, and it is almost a complete gap for $\theta_h = \pi/2$. This applies to all positions in the system, both inside the F-region and inside the S-region. As the energy gap grows with $\theta_h$, this indicates that the fraction of singlet states grows, and that superconductivity is thus being strengthened. This is just in accordance with the analytical derivation in Sec. \ref{ssec:pseudospin}, where the existence of long-ranged singlets was predicted.

\subsubsection{Critical temperature}
\begin{figure}[htb!]
	\begin{center}
		\includegraphics[width=1.0\columnwidth]{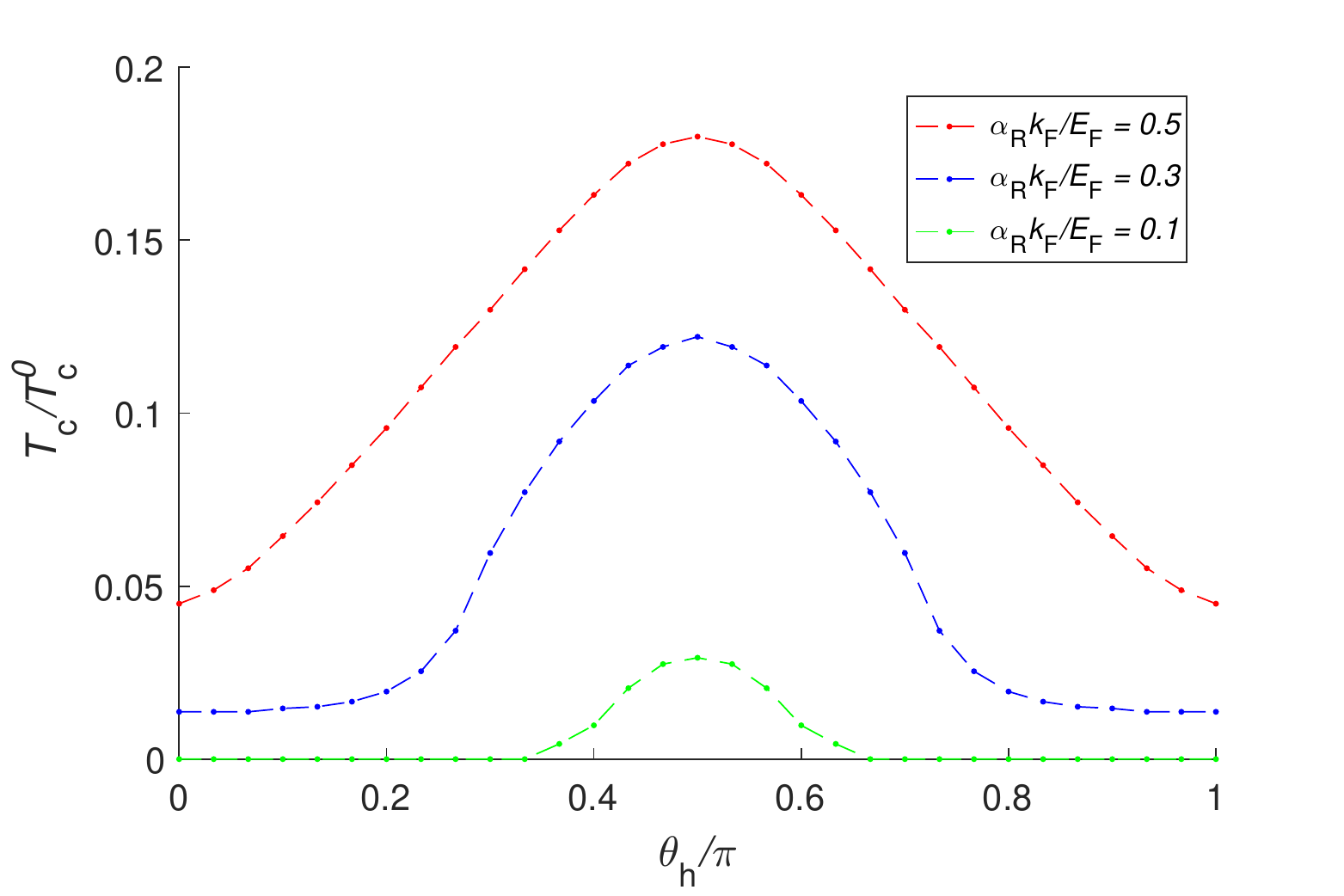}
	\end{center}
	\caption{The critical temperature of an F/S-structure with SOC in the interface plotted for three different Rashba parameters, as function of the magnetization angle, $\theta_h$.}
	\label{fig:TcIntSOCang}
\end{figure}

We have so far seen that the closer $\theta_h$ comes to $\pi/2$, the stronger is the enhancing effect on superconductivity. In order to reveal the exact angular dependence, we have plotted the critical temperature with respect to the magnetization angle in Fig. \ref{fig:TcIntSOCang}. The analysis is done for three different Rashba coupling strengths. The magnetic field is as before, $h_0 / E_{\rm F} = 0.3$.
Firstly, these results confirm that the Hamiltonian is invariant under the transformation $\theta_h \rightarrow \pi - \theta_h$, as the plot is symmetric about $\pi/2$. Furthermore, the results clearly indicate that the closer the magnetization angle is to $\pi/2$, the more robust is the superconducting state. This is an interesting result, as we are able to control the critical temperature by adjusting a macroscopic parameter. Although not directly comparable, these results show similar behaviour as obtained by a quasiclassical approach in the diffusive limit in Ref. \citen{Jacobsen2015, ouassou_scirep_16}. In these works, it was shown that for equal weights of Rashba and Dresselhaus SOC, rotating the magnetic field over an interval of $\pi/2$ causes the critical temperature to go from minimum to maximum. This is just what we found for the F/S-structure with interfacial SOC studied here, with a fully quantum mechanical approach.

In summary, the F/S-structure with interfacial SOC shows interesting properties. Firstly, it allows for controlling the critical temperature by adjusting macroscopic factors such as the magnetic field. Secondly, it may be used to control triplet amplitudes. Such a structure therefore serves as a promising alternative to magnetic multilayers for the purpose of achieving superconducting spin-valve effects.

\subsection{In-plane SOC in an S/F/S-structure}
\label{ssec:InplaneSOC_SFS}

Our observations from the previous section, and the predictions made in Sec. \ref{ssec:pseudospin}, motivates us to look at the case of in-plane SOC. That is, the combined presence of SOC and magnetic fields gives rise to long-ranged singlet pairs. The analysis so far predicts that the closer these interactions are in space, the more prominent the effects will be. In-plane SOC inside a ferromagnet maximizes the spatial copresence of the spin-orbit interaction and the exchange field's effect on the electrons, and we thus expect a larger relative amount of long-ranged singlet pairs as compared to with interfacial SOC. We start out by looking at a S/F/S-structure, with in-plane SOC in the F-region. One way to realize such a setup, a so-called Rashba ferromagnet, is to use a thin film of a strong transition metal ferromagnet, \eg Fe, Co or Ni. The Rashba effect could be further enhanced by adding a thin layer ($\sim 1$ nm) of a heavy metal. This setup mimics a situation where SOC and ferromagnetic order coexist, albeit not entirely homogeneously.

\begin{figure}[htb!]
\centering
\includegraphics[width=1.0\columnwidth]{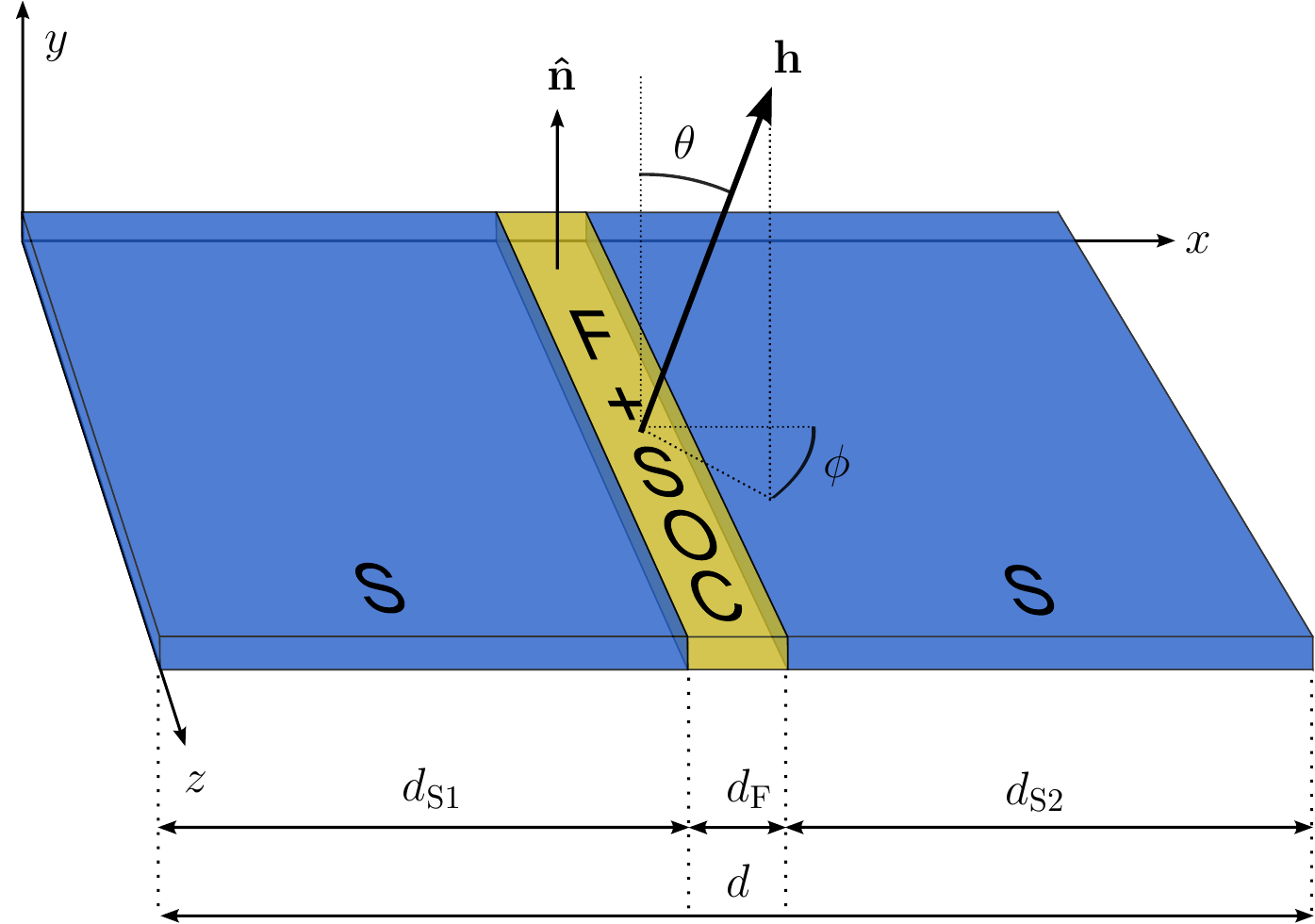}
\caption{An illustration of the S/F/S-structure with in-plane SOC in the F-region. The system considered is in reality not of restricted length along the $z$-axis, but is of infinite extent in this direction. Moreover, the structure is of zero height, that is of no extent in the $y$-direction.}
\label{SFS_sketch}
\end{figure}

As SOC in general is expected to protect a fraction of the singlets, we need not define a very long system for superconductivity to sustain. Therefore, the full system length is only defined to be $1.1 \xi_0$. The S-regions are of length $d_{\rm S1}/\xi_0 = d_{\rm S2}/\xi_0 = 0.5$, which leaves the F-region with SOC of length $d_{\rm F}/\xi_0 = 0.1$ between the S-regions. The system is illustrated in Fig. \ref{SFS_sketch}. There is no applied phase-difference between the superconductors.
As opposed to in the case of interfacial SOC, where a spin-rotational symmetry remained in the system, the spin-rotational symmetry is completely broken by the combined effect of the magnetic field and SOC. We have to keep the magnetic field completely general, and write it as

\begin{equation}
\ve{h} = h_0 \bigg( \cos(\phi)\sin(\theta)\hat{\ve{x}} + \sin(\phi)\sin(\theta)\hat{\ve{z}} + \cos(\theta) \hat{\ve{y}} \bigg),
\end{equation}

\noindent where $\phi$ is the azimuthal angle and $\theta$ is the polar angle of slightly modified spherical coordinates, that is with $y$ and $z$ having changed roles. Moreover, the SOC Hamiltonian becomes

\begin{equation}
H_{\rm SO} = \alpha_{\rm R}\Big[ k_x \sigma_z - k_z \sigma_x \Big] + \frac{\alpha_{\rm R} \sigma_z}{2i}\Big[ \delta(x - x_{\rm L}) - \delta(x - x_{\rm R})  \Big],
\label{eq:H_SFS}
\end{equation}

\noindent where $x_{\rm L}$ and $x_{\rm R}$ are the $x$-coordinates of the left and right boundaries of the SOC-region respectively, and where the position dependence of the Rashba parameter has been suppressed in the notation. Interestingly, we see that the requirement of rendering the Hamiltonian Hermitian leads to an effective barrier term at the interfaces which looks like a spin-dependent scattering potential with an imaginary amplitude. This term may seem like an unwanted term. It introduces complex numbers on the diagonal of $H_{\rm SO}$, which in general could cause complex eigenvalues, resulting in complex, unphysical energies. However, the actual matrix elements entering the diagonal of the BdG equations in Eq. (9)
remain purely real even in the presence of the additional $\delta$-function term,
as can be verified by direct insertion. For this analysis, we set magnetic field strength to $h_0/E_{\rm F} = 0.1$, and the Rashba coupling strength is set to $\alpha_{\rm R} k_{\rm F}/E_{\rm F} = 0.4$. 

\subsubsection{Pair amplitudes}
\begin{figure}[htb!]
	\begin{center}
		\includegraphics[width=1.0\columnwidth]{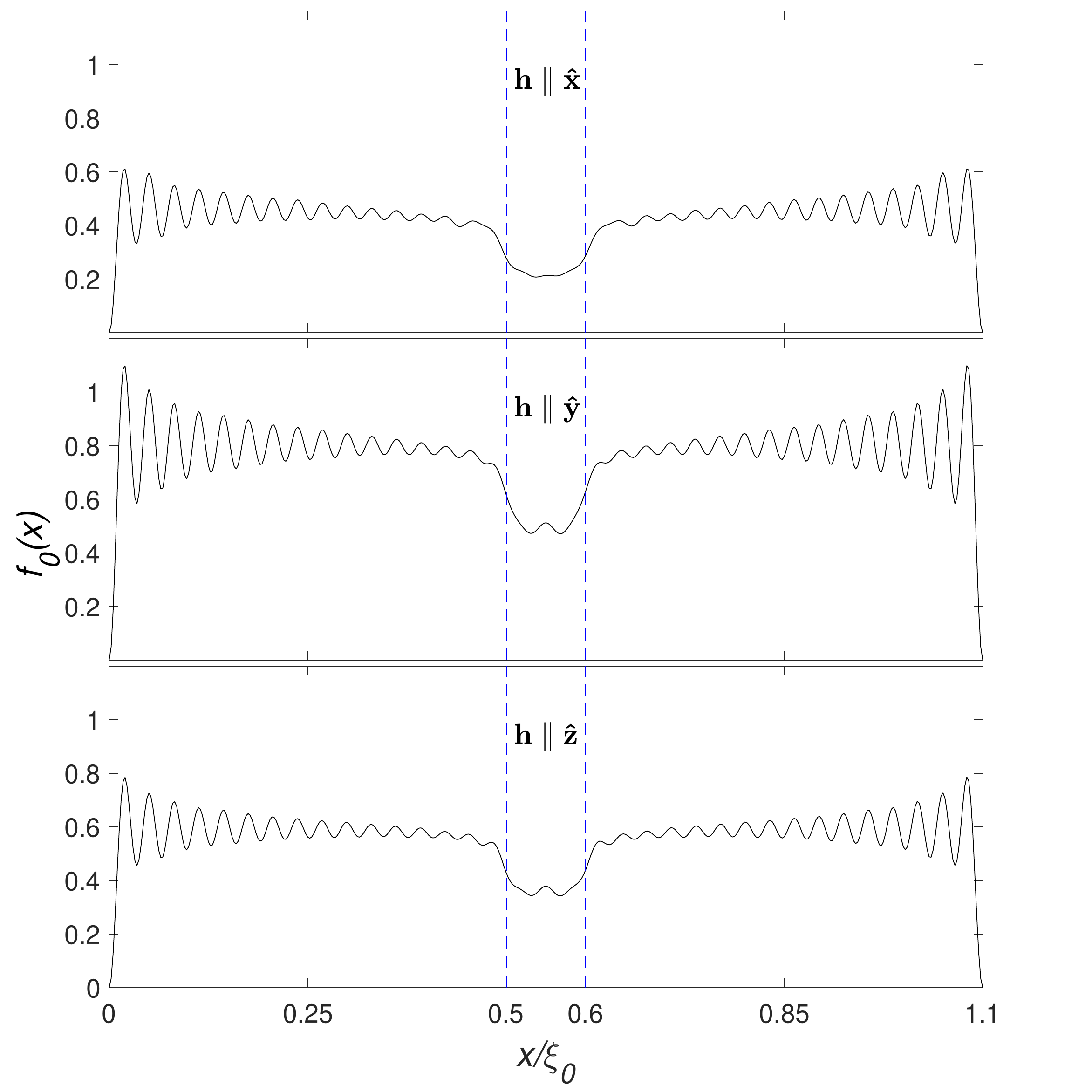}
	\end{center}
	\caption{The singlet amplitude plotted for the S/F/S-structure with in-plane SOC in the F-region, for magnetization along the $x$-, $y$- and $z$-axis. The dotted blue lines indicate the junctions between the F- and S-regions. The Rashba parameter has been set to $\alpha_{\rm R} k_{\rm F} / E_{\rm F} = 0.4$.}
	\label{F_SFS_SO}
\end{figure}

The singlet amplitudes for magnetization along $x$-, $y$- and $z$-axis are plotted in Fig. \ref{F_SFS_SO}.
The qualitative behaviour of the singlet amplitudes in these magnetization setups are all approximately the same. There is however a significant quantitative difference between the different setups. The superconducting state seem to prefer the $y$-alignment of the magnetic field, and is most suppressed by an $x$-aligned field. Hence, as with interfacial SOC, in-plane SOC introduces a prominent dependence upon the direction of the magnetic field. If SOC was switched off, the singlet amplitudes would drop to zero no matter the magnetization direction, implying that SOC once again shows an enhancing effect on superconductivity.

\begin{figure}[htb!]
\centering
\includegraphics[width=1.0\columnwidth]{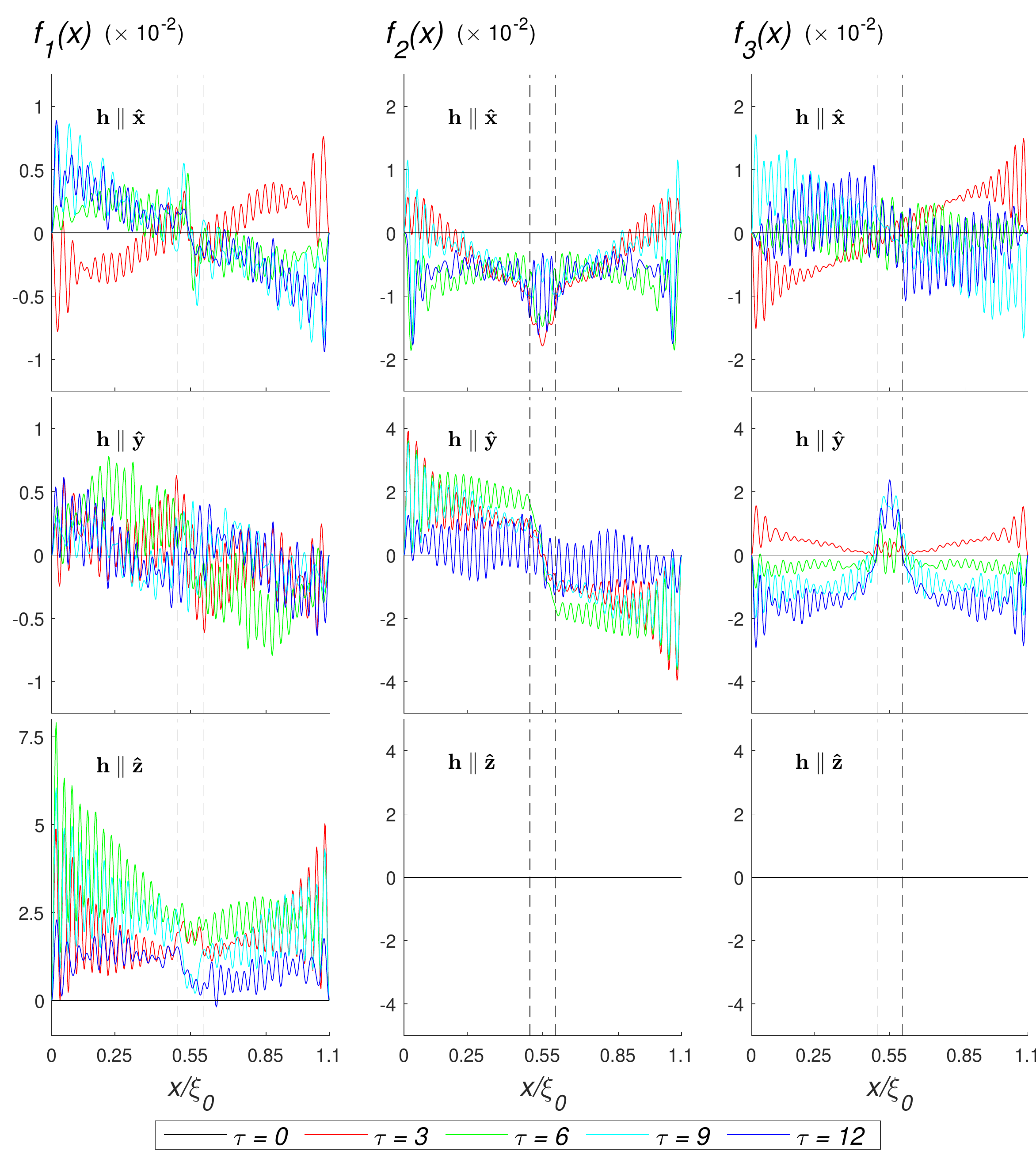}
	\caption{The triplet amplitudes for the S/F/S-structure with in-plane SOC in the F-region for magnetization along the $x$-, $y$- and $z$-axis. The results are plotted for five different relative times $\tau$, as indicated by the legend. The black dotted lines indicate the junctions between the different regions. Note that the axes are scaled differently, and the graphical amplitudes are thus not directly comparable.}
	\label{Triplet_SFS_SO}
\end{figure}

The triplet amplitudes for the same magnetization setups are plotted in Fig. \ref{Triplet_SFS_SO}. Note that the axes are scaled differently, and the graphical amplitudes are thus not directly comparable between the different plots.
\noindent If SOC was switched off, the rotation of the magnetic field would only cause the triplet amplitudes to rotate between each other. An $x$-aligned, $y$-aligned and $z$-aligned magnetization would have given a non-zero $f_2$-amplitude, $f_3$-amplitude and $f_1$-amplitude, respectively. For each magnetization configuration, all other than the mentioned triplet amplitude would have been identically zero. With SOC switched on however, all triplet amplitudes appear for the $x$- and $y$-aligned fields, while only the $f_1$-amplitude remains non-zero for the $z$-aligned field.

There are two reasons to the appearance of other triplet amplitudes. Firstly, as explained in Sec. \ref{ssec:pseudospin}, SOC induces magnetic impurities in the junction between the S- and F-regions. These cause an inhomogeneous magnetization configuration when the magnetic field points along the $x$- or $y$-axis, which alone would result in two triplet amplitudes to appear. Secondly, SOC introduces spin-mixing when the magnetic field is not either orthogonal or parallel to the SOC-induced field. These effects combined generally cause all triplet amplitudes to be present, except for $\ve{h} \parallel \zh$, where no spin mixing occurs and thus only one non-zero triplet amplitude appears.

\subsubsection{LDOS}
\begin{figure}[htb!]
	\begin{center}
		\includegraphics[width=1.0\columnwidth]{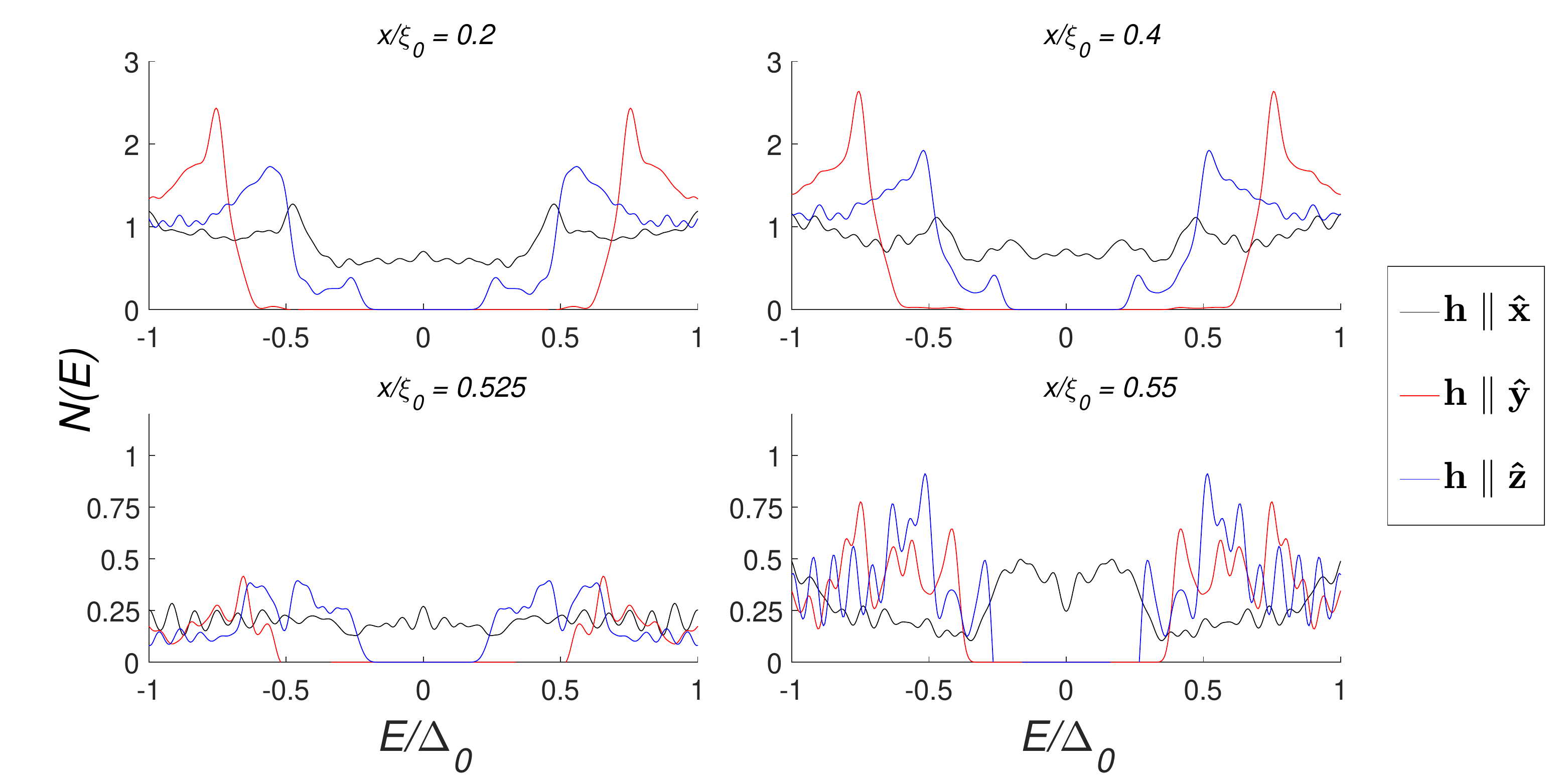}
	\end{center}
	\caption{The LDOS at four different positions inside the S/F/S-structure with in-plane SOC in the F-region. The positions are indicated in the plots, and the colour coding is indicated by the legend. The results are obtained with $N_{\perp}$ = 2000.}
	\label{LDOS_SFS_SO}
\end{figure}

As magnetization in either the $x$-, $y$- or $z$-directions clearly give different pair amplitudes, it makes an interesting analysis to take a closer look at the configuration of states around the Fermi energy for each case. The LDOS at four different positions have therefore been plotted in Fig. \ref{LDOS_SFS_SO}. The upper two plots show the density of states at two different positions inside the left S-region, while the two lower plots do the same for inside the F-region.
\noindent Inside the S-region, there is a fully developed energy gap for magnetization in the $y$- and $z$-directions, with $\ve{h} \parallel \hat{\ve{y}}$ giving the largest gap. The gap is much less developed for the $x$-aligned magnetic field. Inside the F-region, the amplitudes are being suppressed for all system setups, with an average of about $0.25$ outside the band gap region. This is both an effect of the magnetic field, which suppresses certain spin-configurations, as well as due to SOC suppressing states dependent upon both their momentum and spin.

The band gap is fully developed at both positions inside the F-region for both the $y$- and $z$-aligned fields. Once again, the gap is widest for the $y$-aligned field. These results are consistent to the results obtained for the singlet amplitudes. In general, the band gap seems to be wider and more prominent for $\ve{h} \parallel \hat{\ve{y}}$, and weakens for $\ve{h} \parallel \hat{\ve{z}}$ and $\ve{h} \parallel \hat{\ve{x}}$, in that order.

\subsubsection{Critical temperature}
\begin{figure}[htb!]
	\begin{center}
		\includegraphics[width=1.0\columnwidth]{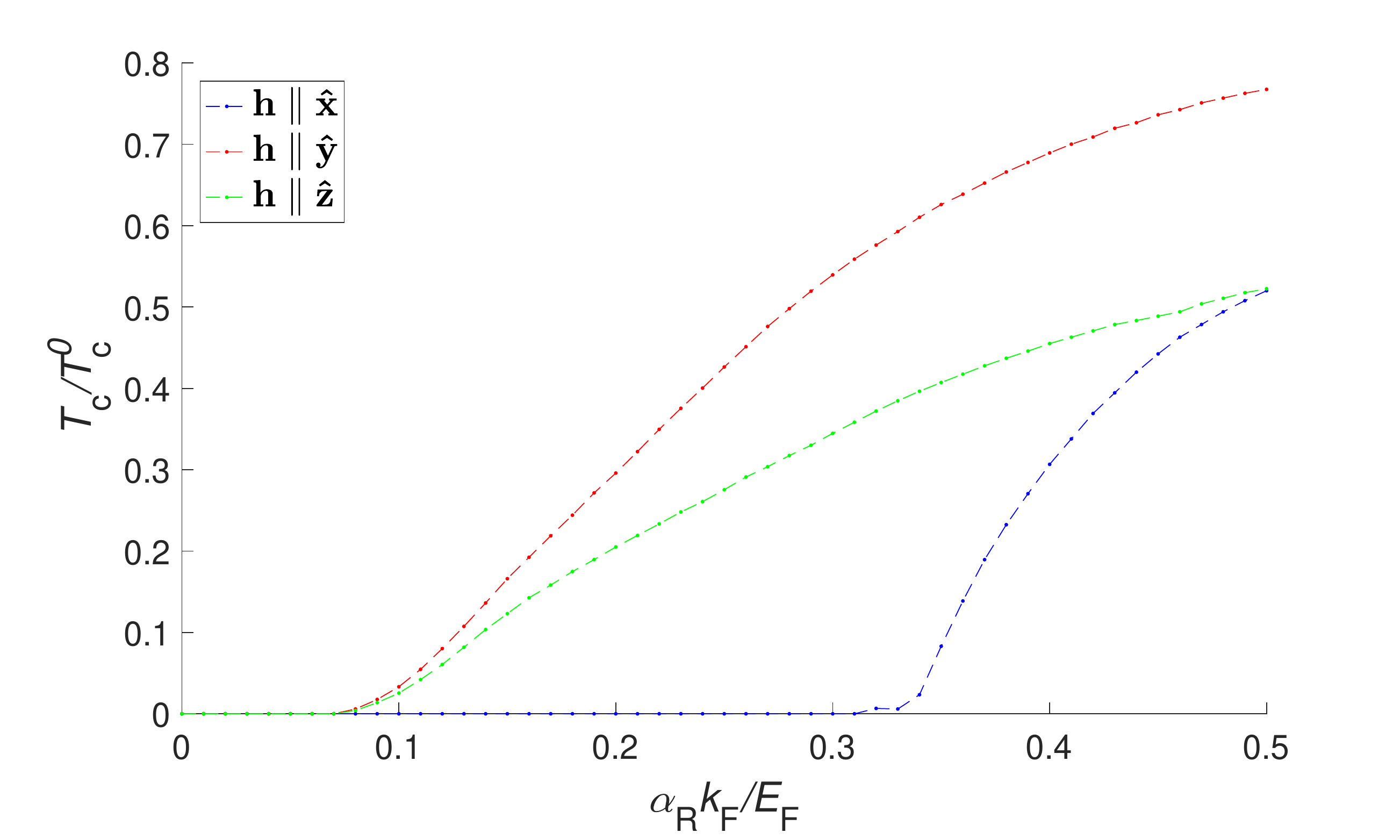}
	\end{center}
	\caption{The critical temperature plotted with respect to the Rashba parameter for the S/F/S-structure with in-plane SOC in the F-region. The magnetic field strength is set to $h_0/E_{\rm F} = 0.1$. Each line represents a magnetic field orientation along an axis, orthogonal to the others, as indicated by the legend. }
	\label{Tc_SFS_SO_alpha}
\end{figure}

The analysis so far has been performed with a constant Rashba parameter. We follow up this analysis by investigating how varying the Rashba parameter affects the physics of the system. This analysis is once again performed for the magnetic field pointing in both the $x$-, $y$- and $z$-directions. The results are plotted in Fig. \ref{Tc_SFS_SO_alpha}.

The critical temperature increases with increasing $\alpha_{\rm R}$ for all three magnetic field configurations. The degree to which the temperature rises differ between all configurations, as we should expect after the analysis so far. The critical temperature is generally higher for a magnetic field pointing in the $y$-direction. For magnetic field configurations in the $xz$-plane, that is in the plane which is spanned by the physical system, the critical temperature is different for small Rashba parameters. The critical temperature is in general higher when the magnetic field is pointing in the $z$-direction, but this difference vanishes almost entirely as ${\alpha_{\rm R} k_{\rm F}}$ approaches $0.5 E_{\rm F}$. These results are consistent with what observed for the singlet amplitude and for the LDOS. However, this analysis also brings some new and interesting observations, which can help us understand the physics better. 

We start by looking into the easily visible difference between magnetization along the $y$-direction and in the $xz$-plane. It is obvious from Fig. \ref{Tc_SFS_SO_alpha} that superconductivity is most resistant to thermal effects with a magnetic field along the $y$-axis. This result is rather straightforward to understand. The superconductivity-enhancing effect caused by interfacial SOC was observed to be most prominent at $\theta_h = \pi/2$, as is also predicted by theory. For the S/F/S-system considered here, the SOC-induced fields are always parallel to the $xz$-plane, perpendicular to the $y$-axis. Thus, for magnetization in the $y$-direction, the requirement for maximal effect of SOC is always satisfied, which explains the higher $T_{\rm c}$.

In a pure F/F/S-structure, a perpendicular relative orientation of neighbouring magnetic field regions causes a lower critical temperature than a parallel alignment \cite{Wu2012}. This is due to the long-range triplet production, which effectively causes another channel of triplet leakage to occur. We can use this result to explain the $\phi$-dependence of $T_{\rm c}$, that is the difference between the $x$- and $z$-directions. When $\phi = \pi/2$, all magnetic fields in the system are either parallel or antiparallel. Hence, in this configuration, only the short-range triplet channel is open. When $\phi$ is decreased however, the production of long-range triplet pairs is increased. This effect reaches its maximum at $\phi = 0$. Another channel of leakage is thus opened by the magnetic field configuration when $0 \le \phi < \pi/2$, which implies lower critical temperature for an $x$-aligned magnetic field than for a $z$-aligned field. For an increasing Rashba parameter however, the amount of short-ranged triplets are reduced, which further implies less leakage into other triplet channels. Thus for increasing $\alpha_{\rm R}$, $T_{\rm c}$ should be less sensitive to changes in $\phi$. As is evident from the results, full $\phi$-invariance seems to occur at $\alpha_{\rm R} k_{\rm F}/E_{\rm F} \approx 0.5$.

\begin{figure}[htb!]
	\begin{center}
		\includegraphics[width=1.0\columnwidth]{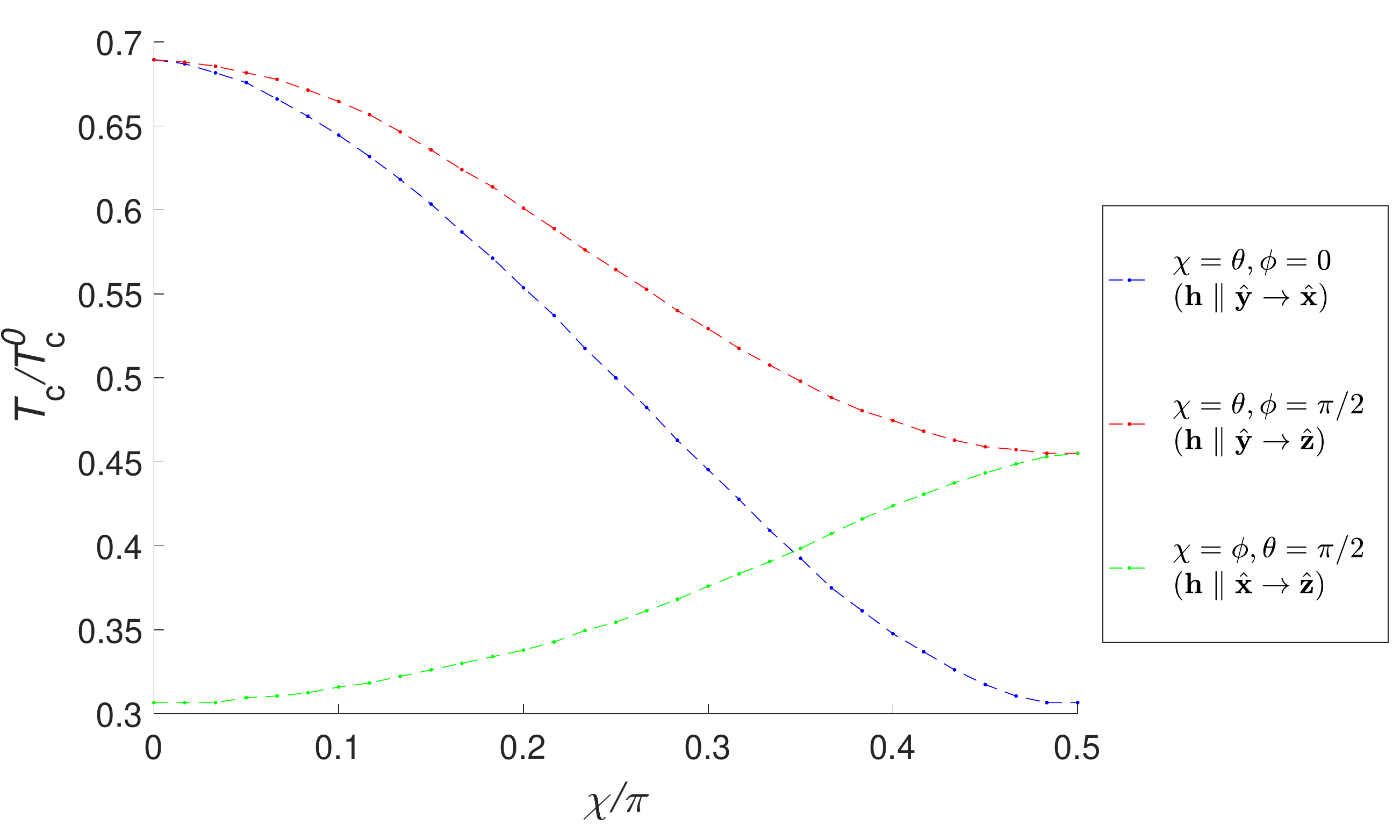}
	\end{center}
	\caption{The critical temperature in the S/F/S-structure with in-plance SOC plotted with respect to different magnetization angles, with $\alpha_{\rm R} k_{\rm F}/E_{\rm F} = 0.4$. $\theta$ and $\phi$ have been rotated between $0$ and $\pi/2$ in the $xy$-plane (blue), $zy$-plane (red) and $zx$-plane (green). The angle $\chi$ represents either $\phi$ or $\theta$, and is specified by the legend for each individual line.}
	\label{Tc_SFS_SO_rot}
\end{figure}

In order to make the analysis complete, and in order to reveal the exact angular dependence, the critical temperature as function of the magnetization angles $(\phi,\theta)$ is plotted in Fig. \ref{Tc_SFS_SO_rot}. The plot contains three graphs, each of which corresponds to rotation in either the $xy$-, $yz$- or $zx$-plane. The Rashba parameter has been set to $\alpha_{\rm R} k_{\rm F} / E_{\rm F} = 0.4$.
The results are consistent with the analysis made in the discussion of magnetization in the $x$-, $y$- or $z$-direction. We also observe that the graphs are all strictly increasing or decreasing, and contain thus no local minima or maxima. The transition between the different extrema, namely magnetization along the coordinate axes, happens smoothly. There are no intermediate angles at which effects other than those discussed up until now occur.

Fig. \ref{Tc_SFS_SO_rot} shows that the largest change in $T_{\rm c}$ by rotating the magnetic field happens for rotation in the $xy$-plane. The difference between $\theta = 0$ and $\theta = \pi/2$ is almost $0.4 T_{\rm c}^{0}$. This structure thus has a great potential in controlling $T_{\rm c}$ by adjusting both the SOC-strength and the magnetization angles. It also serves a candidate for controlling the triplet production, as magnetization along the $x$-, $y$- and $z$-axis all give different properties for the triplet amplitudes. Additionally, these effects are obtainable for a structure of only $1.1\xi_0$, which is generally shorter than required for clean ferromagnet-superconductor-structures.

\subsection{In-plane SOC in an S/F-structure}
\label{ssec:InplaneSOC_SF}

\begin{figure}[htb!]
\centering
\includegraphics[width=1.0\columnwidth]{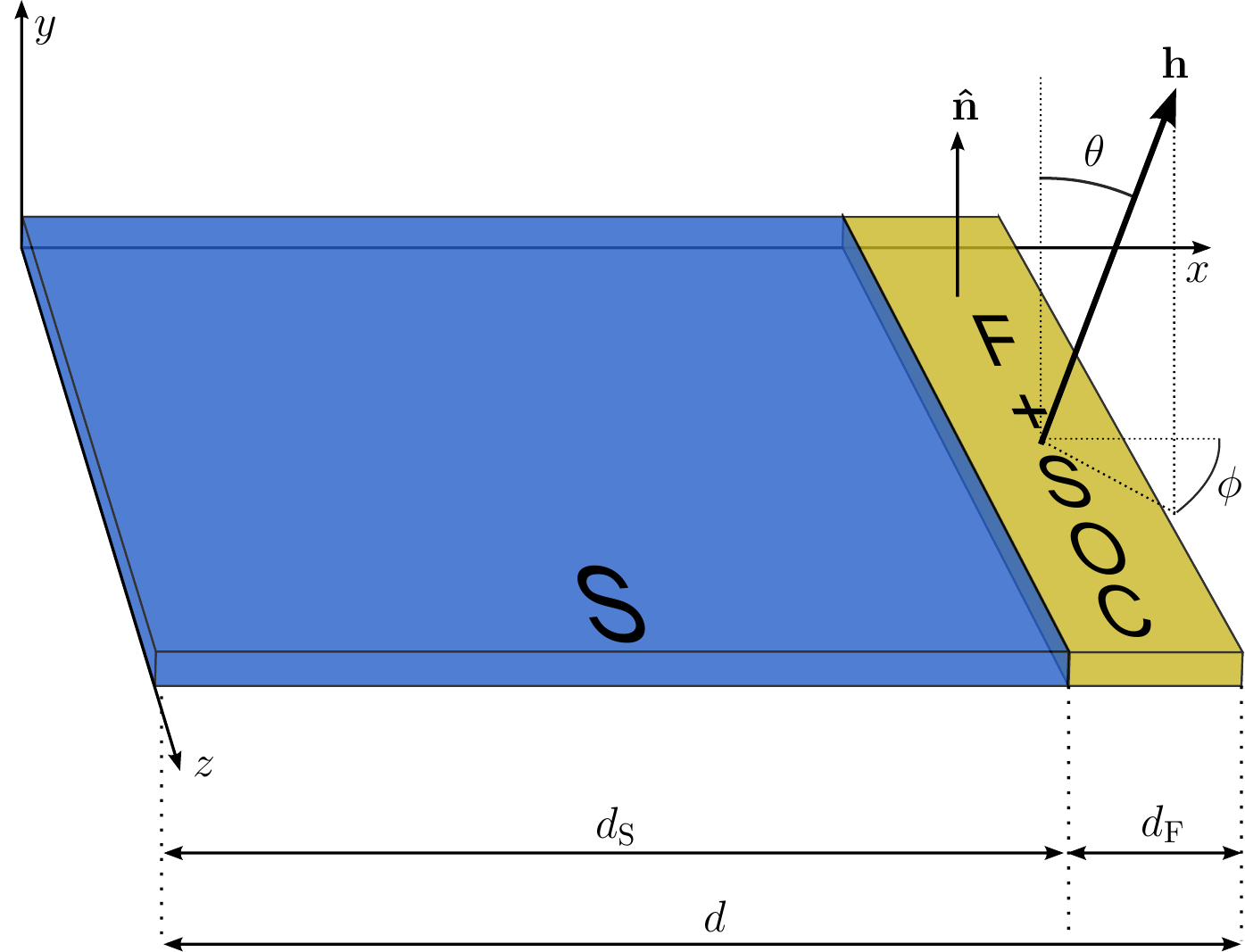}
\caption{An illustration of the S/F-structure with in-plane SOC in the F-region. The system considered is in reality not of restricted length along the $z$-axis, but is of infinite extent in this direction. Moreover, the structure is of zero height, that is of no extent in the $y$-direction.}
\label{SF_sketch}
\end{figure}

\noindent
We will now look at in-plane SOC in an S/F-structure. The structure will have equal dimensions as the previous S/F/S-structure, only with the F-region to the far right side of the system. That is, the S-region is of length $d_{\rm S} = \xi_0$, while the F-region is of length $d_{\rm F} = 0.1 \xi_0$. We still use $\alpha_{\rm R} k_{\rm F}/E_{\rm F} = 0.4$ and $h_0 / E_{\rm F} = 0.1$. The system is illustrated in Fig. \ref{SF_sketch}.
The qualitative difference between this structure and the S/F/S-structure is that this is a bilayer structure rather than a trilayer, and that the SOC-region now forms a boundary region. 
The results are very similar to the S/F/S-structure, and we will therefore not give a complete treatment of this structure. We will restrict the analysis to include the critical temperature plots analogous to the ones given for the S/F/S-structure, and the discussion will mainly focus on the differences.

\subsubsection{Critical temperature}

\begin{figure}[htb!]
	\begin{center}
		\includegraphics[width=1.0\columnwidth]{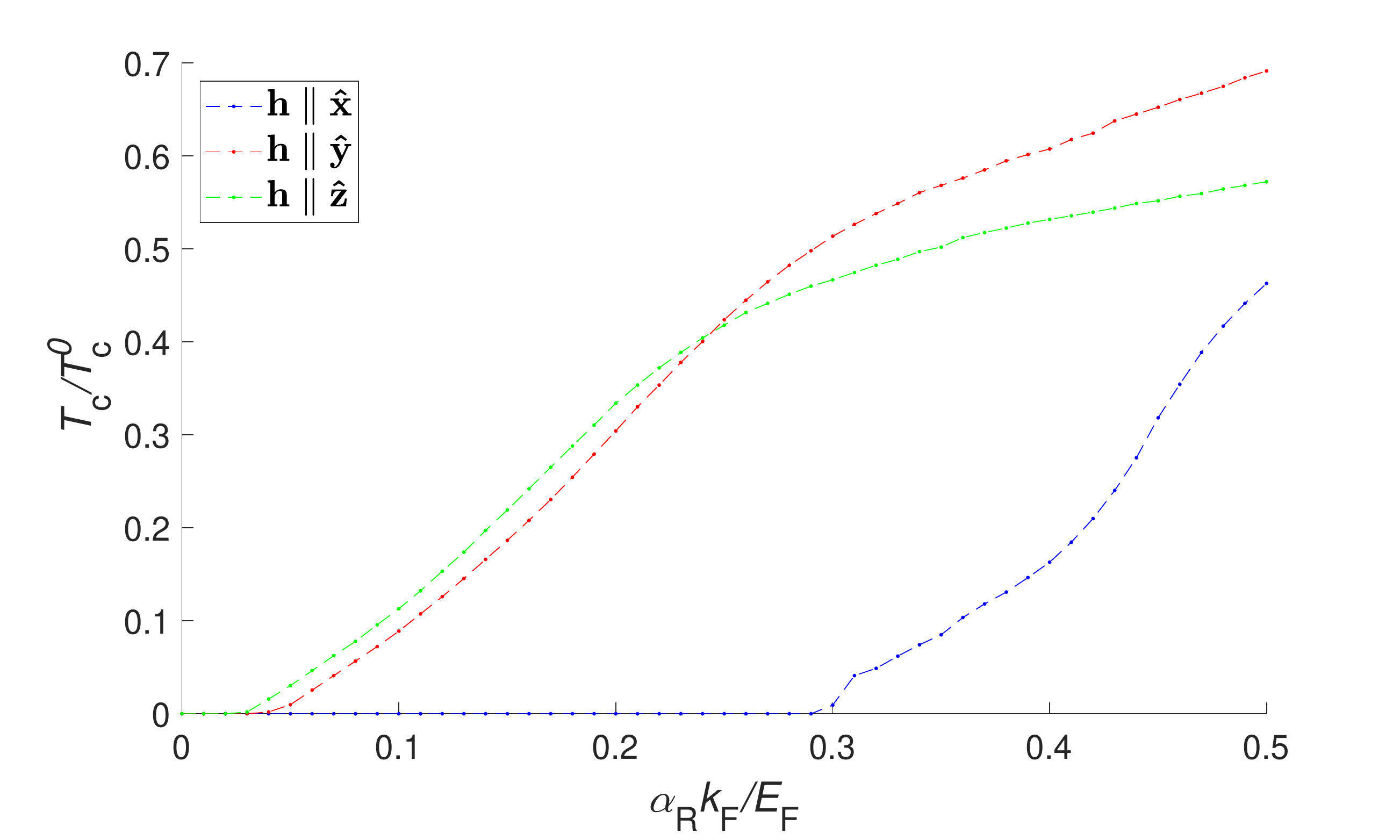}
	\end{center}
	\caption{The critical temperature plotted with respect to the Rashba parameter for the S/F-structure with in-plane SOC in the F-region. The magnetic field strength is set to $h_0/E_{\rm F} = 0.1$. Each line represents a magnetic field orientation along an axis, orthogonal to the others, as indicated by the legend.}	\label{Tc_SF_SO_alpha}
\end{figure}

The critical temperature with respect to the Rashba coupling strength, $\alpha_{\rm R}$, is plotted in Fig. \ref{Tc_SF_SO_alpha}. There are several qualitative similarities to the corresponding plot for the S/F/S-structure, given in Fig. \ref{Tc_SFS_SO_alpha}. Firstly, we observe that the $x$-aligned magnetic field has a clearly visible suppressed critical temperature compared to with the magnetic field pointing in either the $y$- or $z$-directions. With an increasing Rashba parameter, we also observe that the critical temperature is strictly increasing for all of the three magnetization configurations.
The most interesting observations might however be the differences. We see that for all configurations, the critical temperature is non-zero at a lower Rashba parameter than in the S/F/S-structure. This may be explained from the fact that the superconductor length here is twice the size of the lengths of the two S-regions in the S/F/S-structure. With no SOC, few singlet pairs may tunnel through the F-region, and the two regions are in that sense more or less decoupled. In the S/F-structure in comparison, the doubled length of the S-region makes superconductivity arise at an earlier stage. As we increase the Rashba coupling strength however, the critical temperature of the S/F/S-system eventually passes that of the S/F-system. At this point, more pairs may pass through the F-region in the S/F/S-structure, and we might say that the F-region gradually adapts normal metal properties. While in the S/F-structure, the F-region is at the boundary, and the singlet pairs cannot simply tunnel through into a new S-region, but rather has to be reflected and pass through the F-region once more before re-entering the S-region. This results in more triplet conversion, which explains why the critical temperature of the S/F/S-region eventually becomes larger than that of the S/F-region.

\begin{figure}[htb!]
	\begin{center}
		\includegraphics[width=1.0\columnwidth]{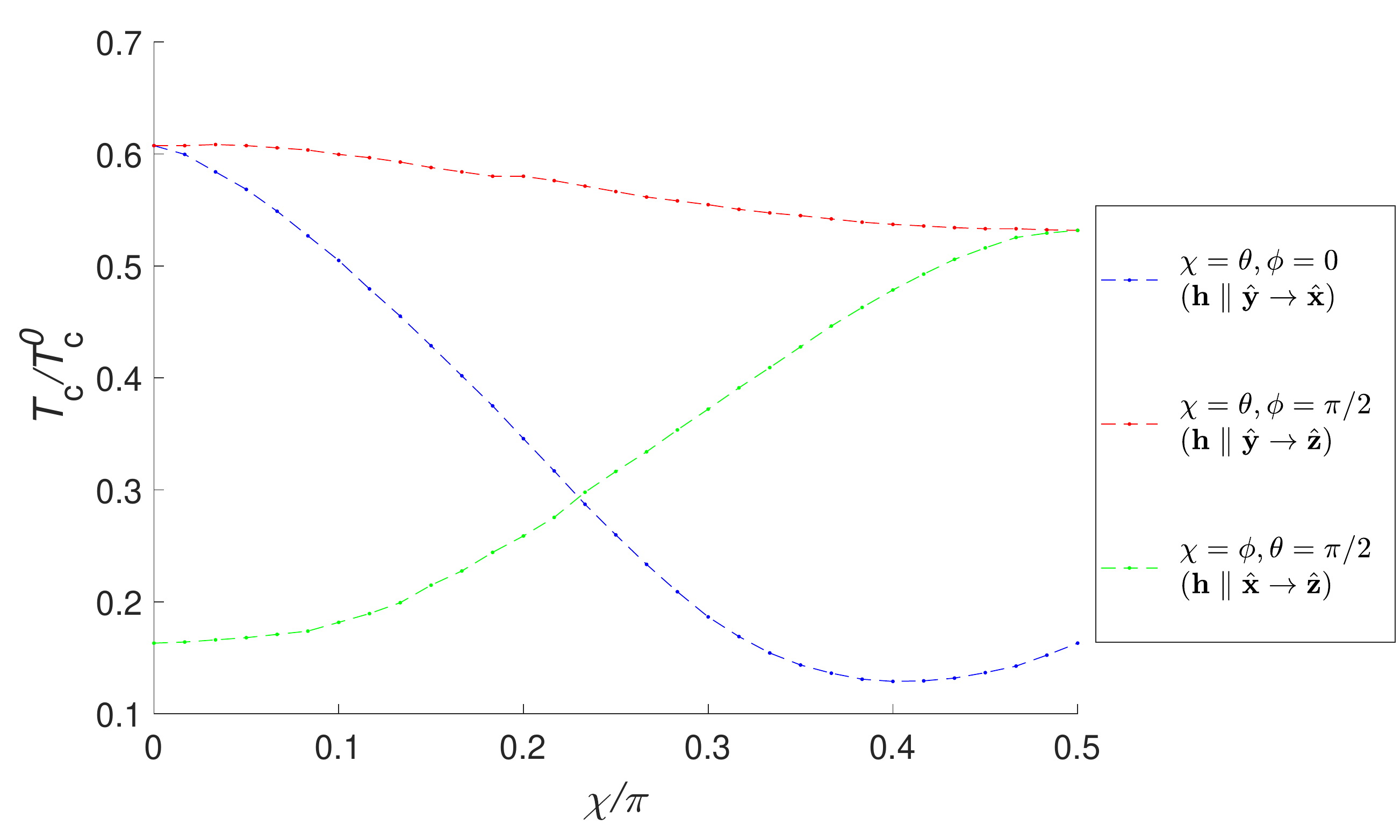}
	\end{center}
	\caption{The critical temperature in the S/F-structure with in-plance SOC plotted with respect to different magnetization angles, with $\alpha_{\rm R} k_{\rm F}/E_{\rm F} = 0.4$. $\theta$ and $\phi$ have been rotated between $0$ and $\pi/2$ in the $xy$-plane (blue), $zy$-plane (red) and $zx$-plane (green). The angle $\chi$ represents either $\phi$ or $\theta$, and is specified by the legend for each individual line.}
	\label{Tc_SF_SO_rot}
\end{figure}

In Fig. \ref{Tc_SF_SO_rot}, the critical temperature is plotted with respect to the magnetization angles $(\phi,\theta)$ in the $xy$- $xz$- and $yz$-plane. This should be compared to the results for the S/F/S-structure plotted in Fig. \ref{Tc_SFS_SO_rot}.
We observe that for the rotations in the $xz$- and $yz$-plane, the local extrema of $T_{\rm c}$ seem generally not to be where the magnetization is along one of the coordinate axes, but rather shifted somewhat from this point. This is interestingly just what we observe when rotating the relative magnetization direction in a clean F/F/S-structure. In the S/F/S-structure with SOC, there was no such effect. This is likely due to the fact this effect occurs on both sides of the F-region in the S/F/S-structure, effectively adding together to produce results where the extrema are found at $\phi$ and $\theta$ equal to 0 and $\pi/2$. The S/F-structure with in-plane SOC may therefore, at least to some extent, serve the same purpose as an F/F/S-structure. Put in other words, the F-region with in-plane SOC may to some extent serve as a substitute for an F/F-region.


\section{Conclusion}

The effects of strong Rashba spin-orbit coupling in ferromagnet-superconductors-structures have been analyzed with both an analytical and a numerical approach. We first did a theoretical analysis in which the $s$-wave spin-singlet state was projected onto the pseudospin eigenbasis in a system where a spin-orbit field and an exchange field coexist throughout the entire non-superconducting material. The analysis showed that the spin-singlet state is projected onto the eigenbasis as a linear combination of a of a short-ranged pseudospin-singlet state and a long-ranged pseudospin-triplet state, depending upon both the relative orientation and the strengths of the exchange and spin-orbit fields. The spin-singlet therefore gains a long-ranged component, which can traverse through the system with a slow decay. The theoretical analysis predicts that the copresence of spin-orbit coupling and magnetic fields can raise the critical temperature compared to if spin-orbit coupling is absent.

The numerical calculations support the predictions of the theoretical analysis. We explored interfacial spin-orbit coupling between a ferromagnet and a superconductor, as well as in-plane spin-orbit coupling in the ferromagnetic region of an S/F/S- and an S/F-structure. Both the pair amplitudes, local density of states and the critical temperature showed to be strongly dependent upon the direction of the exchange field. By rotating the relative orientation between the spin-orbit coupling and exchange fields, or by adjusting either the magnetic field or Rashba coupling, one may therefore control the superconducting properties of the system, making these structures possible candidates for use in cryogenic spintronics components.

\acknowledgments
M. Amundsen, N. Banerjee, S. Jacobsen, J. A. Ouassou, and V. Rising{\aa}rd are thanked for useful discussions. We thank in particular K. Halterman for useful correspondence. J.L. acknowledges funding via the Outstanding Academic
 Fellows program at NTNU, the NV-Faculty, and the Research
 Council of Norway Grant numbers 216700 and 240806.
This work was partially supported by the Research Council of Norway through the funding of the Center of Excellence "QuSpin" project no. 262633.

\appendix

\section{Finding $T_c$ with perturbation theory}
\label{sec:App_Tc}

\noindent We start by defining particle-/hole-amplitude vectors

\begin{align}
\begin{split}
\ve{u}_n(x) =
\begin{pmatrix}
u_{n,\uparrow}(x) \\
u_{n,\downarrow}(x)
\end{pmatrix}
\: ,
\end{split}
\begin{split}
\ve{v}_n(x) =
\begin{pmatrix}
v_{n,\uparrow}(x) \\
v_{n,\downarrow}(x)
\end{pmatrix}
\end{split}
\: ,
\end{align}

\noindent and the matrices

\begin{align}
\begin{split}
\bar{H}_e =
\begin{pmatrix}
H_e & 0 \\
0 & H_e 
\end{pmatrix}
\end{split}
\: ,
\begin{split}
\bar{\Delta} = \textbf{J}_2 \Delta
\end{split}
\: ,
\end{align}

\noindent where $\textbf{J}_2$ is the $(2 \times 2)$ exchange matrix, sometimes also referred to as the backward identity matrix,

\begin{equation}
\textbf{J}_2 = 
\begin{pmatrix}
0 & 1 \\
1 & 0 
\end{pmatrix}
.
\label{eq:Exchange_matrix}
\end{equation}

\noindent This matrix will also be of use later to express cross-coupling terms like $u_{n,\sigma} v_{n,-\sigma}$. Furthermore define $\boldsymbol{\sigma}$ as the vector of Pauli matrices, and a slightly altered vector of Pauli matrices $\tilde{\boldsymbol{\sigma}} = [-\sigma_x, \sigma_y, \sigma_z ]$. By using this notation, the BdG equations take the form

\begin{equation}
\begin{pmatrix}
\bar{H}_e -\ve{h} \cdotp \boldsymbol{\sigma} -\ve{h}_{\rm SO} \cdotp \boldsymbol{\sigma} & \bar{\Delta} \\
\bar{\Delta}^{*} & -\left[\bar{H}_e - \ve{h} \cdotp \tilde{\boldsymbol{\sigma}} + \ve{h}_{\rm SO}\cdotp \tilde{\boldsymbol{\sigma}}\right] 
\end{pmatrix}
\begin{pmatrix}
\ve{u}_n \\
\ve{v}_n
\end{pmatrix}
=
E_n
\begin{pmatrix}
\ve{u}_n \\
\ve{v}_n
\end{pmatrix}
,
\label{eq:BdG_new_not}
\end{equation}

\noindent where we remind ourselves that $\ve{h}_{\rm SO}$ is a momentum-dependent operator, while $\boldsymbol{\sigma}$ and $\tilde{\boldsymbol{\sigma}}$ act in spin space. We now do a perturbation expansion

\begin{align}
\ve{u}_n &= \ve{u}_n^{(0)} \; + \delta \ve{u}_n^{(1)} + \mathcal{O}(\delta^2),
\label{eq:u_pert}
\\
\ve{v}_n &= \ve{v}_n^{(0)} \; + \delta \ve{v}_n^{(1)} + \mathcal{O}(\delta^2),
\label{eq:v_pert}
\\
E_n &= E_n^{(0)} + \delta E_n^{(1)} + \mathcal{O}(\delta^2), \\
\bar{\Delta} &= 0_{\nodag}^{\nodag} \; \; \; + \delta \bar{\Delta}^{(1)} + \mathcal{O}(\delta^2)
\label{eq:Delta_pert}
,
\end{align}

\noindent where $\delta$ is an arbitrary perturbation parameter, which eventually will be set to 1. $\ve{u}_n^{(1)}$ is conventionally assumed to be an orthogonal function to $\ve{u}_n^{(0)}$, that is \\ ${ \int_0^d dx \ve{u}_n^{(1) \dagger}(x) \ve{u}_n^{(0)}(x) = 0}$, and $\ve{v}_n^{(1)}$ is likewise assumed to be orthogonal to $\ve{v}_n^{(0)}$. We have defined the superconducting band gap such that it first enters the equations at order $\mathcal{O}(\delta)$. To zeroth order, Eq. \eqref{eq:BdG_new_not} is diagonal, meaning $\ve{u}_n$ and $\ve{v}_n$ are completely decoupled. This implies that $\ve{u}_n^{(0)}$ and $\ve{v}_n^{(0)}$ have separate energy spectra, $E_n^p$ and $E_n^h$ respectively, where $p$ and $h$ denote particle and hole, and are found by solving the zeroth order BdG equations:

\begin{align}
\Big( \bar{H}_e -\ve{h} \cdotp \boldsymbol{\sigma} -\ve{h}_{\rm SO} \cdotp \boldsymbol{\sigma} \Big) \ve{u}^{(0)}_n(x) &= E^{p}_n \ve{u}^{(0)}_n(x), \\
-\Big( \bar{H}_e -\ve{h} \cdotp \tilde{\boldsymbol{\sigma}} +\ve{h}_{\rm SO} \cdotp \tilde{\boldsymbol{\sigma}} \Big) \ve{v}^{(0)}_n(x) &= E^{h}_n \ve{v}^{(0)}_n(x).
\end{align}

\noindent To order first order, $\mathcal{O}(\delta)$, the BdG equations read

\begin{align}
\Big( \bar{H}_e -\ve{h} \cdotp \boldsymbol{\sigma} -\ve{h}_{\rm SO} \cdotp \boldsymbol{\sigma} \Big) \ve{u}_n^{(1)} + \bar{\Delta}^{(1)} \ve{v}_n^{(0)} &= E_n^{(1)} \ve{u}_n^{(0)} + E_n^{(0)} \ve{u}_n^{(1)},
\label{eq:BdG_1st_u}
\\
-\Big( \bar{H}_e -\ve{h} \cdotp \tilde{\boldsymbol{\sigma}} + \ve{h}_{\rm SO} \cdotp \tilde{\boldsymbol{\sigma}} \Big) \ve{v}_n^{(1)} + \bar{\Delta}^{(1) *} \ve{u}_n^{(0)} &= E_n^{(1)} \ve{v}_n^{(0)} + E_n^{(0)} \ve{v}_n^{(1)}.
\label{eq:BdG_1st_v}
\end{align}

\noindent Now operate on Eq. \eqref{eq:BdG_1st_u} with ${ \sum_{m_{\parallel} \ne n_{\parallel}} \int_0^d dx' \ve{u}_{m}^{(0) \nodag}(x) \ve{u}_{m}^{(0) \dagger}(x')}$, and on Eq. \eqref{eq:BdG_1st_v} with ${ \sum_{m_{\parallel} \ne n_{\parallel}} \int_0^d dx' \ve{v}_{m}^{(0) \nodag}(x) \ve{v}_{m}^{(0) \dagger} (x')}$. Use the orthogonality and completeness relations of $\ve{u}_n$ and $\ve{v}_n$, and the following formulas for the first order corrections are then obtained:

\begin{align}
\ve{u}_n^{(1)}(x) &= \sum_{m_{\parallel} \ne n_{\parallel}} \frac{\int_0^d dx' \ve{u}_m^{(0) \dagger}(x') \bar{\Delta}^{(1)}(x') \ve{v}_n^{(0)}(x') }{E_n^{(0)} - E_m^{p}} \ve{u}_m^{(0)}(x), 
\label{eq:u_1st_order}
\\
\ve{v}_n^{(1)}(x) &= \sum_{m_{\parallel} \ne n_{\parallel}} \frac{\int_0^d dx' \ve{v}_m^{(0) \dagger}(x') \bar{\Delta}^{(1) *}(x') \ve{u}_n^{(0)}(x') }{E_n^{(0)} - E_m^{h}} \ve{v}_m^{(0)}(x)
,
\label{eq:v_1st_order}
\end{align}

\noindent where it is implied in the notation that the perpendicular energy quantum number is equal for all involved wave functions, that is $m_{\perp} = n_{\perp}$. The sum over $m_{\parallel} \ne n_{\parallel}$ is a sum over a complete set of one-dimensional eigenfunctions, with the exception of $m_{\parallel} = n_{\parallel}$, which is not included due to the assumption that the first order corrections are orthogonal to the zeroth order functions. Keep in mind that for the perturbation expansion to be valid, the fractions in Eqs. \eqref{eq:u_1st_order} and \eqref{eq:v_1st_order} have to be $\ll 1$.

We now want to derive an expression for the first order correction to $\Delta(x)$, that is $\Delta^{(1)}(x)$, by using the first order results for the wave functions. First expand $\Delta(x)$ in its Fourier components,

\begin{equation}
\Delta(x) = \sum_q \Delta_q \sin(k_q x),
\label{eq:Delta_expansion}
\end{equation}

\noindent where, as previously, $k_q = q \pi / d$. Equivalently, we may write

\begin{equation}
\Delta_l = \frac{2}{d} \int_0^d dx \Delta(x) \sin(k_l x).
\label{eq:Delta_l}
\end{equation}

\noindent $\Delta(x)$ is non-zero only inside intrinsic superconductors. Thus for a system in the $x$-direction with a non-superconducting region on the interval $[0,x_0)$, and a superconducting material on the interval $[x_0,d]$, we may also write

\begin{equation}
\Delta(x) = \Theta(x-x_0) \Delta(x) = \Theta(x-x_0) \sum_q \Delta_q \sin(k_q x),
\label{eq:Delta_unit_step}
\end{equation} 

\noindent where $\Theta(x-x_0)$ is the unit step function. It seems as though introducing this step function is unnecessary, but it will come of use quite soon. We now insert the definition of $\Delta(x)$, given in Eq. \eqref{eq:Gap}, into Eq. \eqref{eq:Delta_l}, and obtain

\begin{equation}
\Delta_l = \lambda \frac{E_{\rm F}}{2 k_{\rm F} d} \sum_{n} \int_0^d dx \ve{v}_n^{\dagger}(x) \textbf{J}_2 \ve{u}_n^{\nodag}(x) \sin(k_l x) \tanh(E_{n}/2k_{\rm B}T),
\label{eq:Gap_new_notation}
\end{equation}

\noindent where $\textbf{J}_2$ is the exchange matrix defined in Eq. \eqref{eq:Exchange_matrix}. We now do a perturbation expansion of the Fourier coefficients of $\Delta$. Since we are working to first order, we want to find $\Delta_l^{(1)}$, and $\Delta_l^{(0)} = 0$ by assumption. By inserting the perturbation expansion of $\Delta_l$ to first order on the left hand side, and the perturbation expansions of $\ve{u}_n$ and $\ve{v}_n$ in Eqs. \eqref{eq:u_pert} and \eqref{eq:v_pert} on the right hand side of Eq. \eqref{eq:Gap_new_notation}, we obtain

\begin{equation}
\begin{split}
\delta \Delta_l^{(1)} &= \lambda \frac{E_{\rm F}}{2 k_{\rm F} d} \sum_{n} \int_0^d dx \Big( \ve{v}_n^{(0) \dagger}(x) + \delta \ve{v}_n^{(1) \dagger}(x) \Big) \textbf{J}_2 \\
&\Big( \ve{u}_n^{(0)}(x) + \delta \ve{u}_n^{(1)}(x) \Big) \sin(k_l x) \tanh(E_{n}/2k_{\rm B}T).
\label{eq:Gap_1st_order}
\end{split}
\end{equation}

\noindent We observe that there is no term of order $\mathcal{O}(\delta^0)$ on either side of the equation, which is consistent. Now insert the first order results from Eqs. \eqref{eq:u_1st_order} and \eqref{eq:v_1st_order} into \eqref{eq:Gap_1st_order}, expand $\ve{u}_n$ and $\ve{v}_n$ as in Eq. \eqref{eq:Fourier_psi}, and expand $\Delta^{(1)}(x)$ as in Eq. \eqref{eq:Delta_unit_step}. We neglect the terms of order $\mathcal{O}(\delta^2)$ which appear on the right hand side of the equation, and we get the following matrix equation

\begin{equation}
\Delta_l^{(1)} = \sum_k \mathbf{J}_{l k}(T) \Delta_k^{(1)},
\label{eq:T_c_eig2}
\end{equation}

\noindent where the matrix elements $\textbf{J}_{l k}$ are defined by the formula

\begin{equation}
\begin{split}
\textbf{J}_{l k}(T) &= \lambda  \frac{2 E_{\rm F}}{k_{\rm F} d^3}  \sum_n \sum_{m_{\parallel}} \sum_{p,q} K_{pql}\\
&\Bigg\{ 
 \ve{v}_{mq}^{(0) \dagger} \textbf{J}_2 \ve{u}_{np}^{(0)} \frac{\sum_{i,j} \ve{u}_{ni}^{(0) \dagger} \textbf{J}_2 \ve{v}_{mj}^{(0)} K_{ijk} }{E_n^{p} - E_m^{h}} \tanh(\frac{E_n^{p}}{2 k_B T}) \\
&+
\ve{v}_{nq}^{(0) \dagger} \textbf{J}_2 \ve{u}_{mp}^{(0)} \frac{\sum_{i,j} \ve{u}_{mi}^{(0) \dagger} \textbf{J}_2 \ve{v}_{nj}^{(0)} K_{ijk} }{E_n^{h} - E_m^{p}} \tanh(\frac{E_n^{h}}{2 k_B T})
\Bigg\}.
\end{split}
\label{eq:J_matrix2}
\end{equation}


\text{ }\\

\bibliographystyle{apsrev4-1}
\bibliography{ReferencesSOC}

\end{document}